\documentclass[aps,physrev,twocolumn,superscriptaddress]{revtex4-1}

\usepackage[pdftex]{graphicx}
\usepackage{dcolumn}
\usepackage{bm}
\usepackage{amsmath, amssymb,amsfonts}
\usepackage{mathrsfs}
\usepackage{type1cm}
\usepackage[colorlinks,linkcolor=blue,citecolor=blue]{hyperref}

\begin{document}

\title{SU(3) truncated Wigner approximation for strongly interacting Bose gases}

\author{Kazuma Nagao}
\email{knagao@physnet.uni-hamburg.de}
\affiliation{%
Yukawa Institute for Theoretical Physics, Kyoto University, Kitashirakawa Oiwakecho, Sakyo-ku, Kyoto 606-8502, Japan
}%
\affiliation{%
Zentrum f\"ur Optische Quantentechnologien and Institut f\"ur Laserphysik, Universit\"at Hamburg, 22761 Hamburg, Germany\\
}%
\affiliation{%
The Hamburg Center for Ultrafast Imaging, Luruper Chaussee 149, 22761 Hamburg, Germany
}%
\author{Yosuke Takasu}%
\affiliation{%
Department of Physics, Kyoto University, Kitashirakawa Oiwakecho, Sakyo-ku, Kyoto 606-8502, Japan
}%
\author{Yoshiro Takahashi}%
\affiliation{%
Department of Physics, Kyoto University, Kitashirakawa Oiwakecho, Sakyo-ku, Kyoto 606-8502, Japan
}%
\author{Ippei Danshita}%
\affiliation{%
Department of Physics, Kindai University, 3-4-1 Kowakae, Higashi-Osaka, Osaka 577-8502, Japan
}%

\date{\today}

\begin{abstract}

We develop and utilize the SU(3) truncated Wigner approximation (TWA) in order to analyze far-from-equilibrium quantum dynamics of strongly interacting Bose gases in an optical lattice. 
Specifically, we explicitly represent the corresponding Bose-Hubbard model at an arbitrary filling factor with restricted local Hilbert spaces in terms of SU(3) matrices. 
Moreover, we introduce a discrete Wigner sampling technique for the SU(3) TWA and examine its performance as well as that of the SU(3) TWA with the Gaussian approximation for the continuous Wigner function. 
We directly compare outputs of these two approaches with exact computations regarding dynamics of the Bose-Hubbard model at unit filling with a small size and that of a fully connected spin-1 model with a large size. 
We show that both approaches can quantitatively capture quantum dynamics on a timescale of $\hbar/(Jz)$, where $J$ and $z$ denote the hopping energy and the coordination number. 
We apply the two kinds of SU(3) TWA to dynamical spreading of a two-point correlation function of the Bose-Hubbard model on a square lattice with a large system size, which has been measured in recent experiments. 
Noticeable deviations between the theories and experiments indicate that proper inclusion of effects of the spatial inhomogeneity, which is not straightforward in our formulation of the SU(3) TWA, may be necessary.

\end{abstract}

\maketitle

\section{Introduction}

Quantum simulators built with synthetic quantum platforms that are highly controllable have been applied for studying quantum many-body physics in and out of equilibrium. Examples of such quantum simulators include ultracold gases in optical lattices~\cite{Bloch12,Gross17,Trotzky12,Cheneau12,Sandholzer19}, Rydberg atoms in optical tweezer arrays \cite{Browaeys20}, trapped ions~\cite{Lanyon11}, and superconducting circuits~\cite{Ma19,Ye19}. 
Of particular interest is far-from-equilibrium quantum dynamics of isolated many-body systems described by the tight-binding Hubbard-type models, which can be simulated with ultracold gases in optical lattices. 
The quantitative accuracy of such analog quantum simulators for non-equilibrium lattice systems has been examined through direct comparisons with outputs from exact computational methods for some special cases, such as the exact diagonalization for small systems~\cite{Kaufman16} and the matrix-product-state (MPS) approaches for one-dimensional (1D) systems~\cite{Trotzky12,Cheneau12}. 
With the high accuracy confirmed, results obtained from optical-lattice quantum simulators have been exploited in order to test approximate computational methods for quantum many-body dynamics in higher dimensions. 
For instance, it has been shown in Ref.~\cite{Sandholzer19} that the non-equilibrium dynamical mean-field theory can quantitatively capture dynamics of the three-dimensional (3D) Hubbard model subjected to a periodic driving. 
Moreover, in Ref.~\cite{Nagao19}, the Gross-Pitaevskii truncated-Wigner approximation (GPTWA), which is a semiclassical phase-space method on the basis of the GP mean-field theory \cite{Blakie08,Polkovnikov10}, has been directly compared with experimental data regarding dynamics of the 3D Bose-Hubbard model in a weakly interacting regime after a quantum quench. 
It has been shown that the outputs of GPTWA with no free parameter are in good agreement with experimental data for early-time regions. 

In recent years, some experimental works have explored quantum quench dynamics of strongly interacting ultracold gases in two-dimensional (2D) and 3D optical lattices \cite{Braun15,Takasu20}. 
In Ref.~\cite{Takasu20}, an experimental group at Kyoto University has studied sudden-quench dynamics of equal-time single-particle correlation functions for a strongly interacting $^{174}{\rm Yb}$ gas loaded into a deep 2D lattice. 
In contrast to 1D systems, it is generally hard to numerically simulate time evolution of correlation functions in 2D and 3D even on a short timescale. 
It has been found in Ref.~\cite{Takasu20} that the ordinary GPTWA cannot fully capture characteristic properties of the correlation propagation after sudden quenches, e.g., peak and dip properties observed in the correlation signals and saturated values of the correlation at relatively long times. 
This can be attributed to the fact that in the strongly interacting regime the adequate classical limit of the system is not condensates of coherent bosons described by the GP theory. 

In Ref.~\cite{Davidson15}, Davidson and Polkovnikov have introduced a promising phase-space approach for analyzing strongly interacting Bose-Hubbard systems.
This method is called the SU(3) TWA [hereafter SU(3)TWA].
For sufficiently large local interactions, the Bose-Hubbard model reduces to an effective pseudospin-1 model acting on a projected Hilbert space \cite{Huber07,Nagao18}.
In the SU(2) TWA method, which is typically discussed and used in the context of experiments of large-spin systems and arrays of trapped Rydberg atoms~\cite{Schachenmayer15,kunimi2021performance}, this effective model is treated as a Hamiltonian consisting of the SU(2) spin operators for $S=1$~\cite{Polkovnikov10}.
However, for the SU(3) TWA, the model is translated into a Hamiltonian consisting of SU(3) matrices, which gives an alternative phase-space representation of the system with extra five dimensions in addition to the three dimensions of the SU(2) phase space.
Since the local interaction terms of the effective model can be {\it linearized} in the SU(3) matrices, the local particle and hole fluctuations, which produce key effects on the dynamical properties of the strongly interacting regime, are accurately captured at the level of the semiclassical approximation~\cite{Davidson15}.
The TWA method based on the GP trajectories is not suitable to formulate those fluctuations in the strongly interacting limit, just as the Bogoliubov approximation for weakly interacting dilute Bose gases fails to describe the quantum phase transitions to the Mott-insulator phases at low temperatures~\cite{oosten2001quantum}.
We therefore expect that, the SU(3) TWA may simulate the dynamics in the strongly interacting regime of the experiment~\cite{Takasu20}, beyond the capability of the GPTWA, and also the SU(2) TWA. 

In their original work, the performance of the SU(3)TWA was tested by applying it to a fully connected spin-1 model, which has an all-to-all spin-exchange (or hopping) term and can be numerically diagonalized even at a large size. 
However, its quantitative accuracy in realistic cases, where the hopping reaches only nearest neighbors and the system size is large, has not been examined so far.
Furthermore, an effective model that they used to describe Bose-Hubbard systems is valid only for high-filling cases. 
Therefore, their formulation is not directly applicable to unit-filling Bose-Hubbard systems, which are typically considered in the context of the quantum-simulation studies.
We note that a numerical calculation of the SU(3)TWA for a unit-filling experimental setup has been presented in Ref.~\cite{Davidson17Thesis}; however, its explicit formalism has not been provided so far.
 
The goal of this paper is to examine the performance of the SU(3)TWA in simulating quench dynamics of strongly interacting Bose gases in a 2D optical lattice \cite{Takasu20}. 
We extend the previous formalism, which was applied to an effective pseudospin-1 model for the Bose-Hubbard model with large filling factors and strong interactions \cite{Altman02,Nagao16}, to the unit-filling case \cite{Huber07,Nagao18} corresponding to the experimental setup. 
As a technique to evaluate the phase-space integration emerging in the SU(3)TWA, we will employ two different approaches, i.e., the Gaussian approximation for the (continuous) Wigner function \cite{Davidson15} and the discrete TWA (DTWA) approach \cite{Schachenmayer15,Zhu19,kunimi2021performance}. 
In particular, the DTWA approach is thought to be better than the Gaussian approach. 
Indeed, the numerical sampling of the DTWA can be readily carried out without approximation of the probability distribution functions (see also Refs.~\cite{Schachenmayer15,Zhu19}).
In this paper, we also study the performance of a DTWA sampling for the SU(3)TWA via large-scale numerical simulations for a fully connected spin-1 model.
A numerical simulation on the basis of the DTWA scheme will be compared with the experimental data as well as that of the Gaussian approximation.

The remainder of this paper is organized as follows: 
In Sec.~\ref{sec:2}, we introduce an effective pseudospin-1 model for the Bose-Hubbard Hamiltonian in a strongly interacting regime and a fully connected spin-1 model, respectively.
In Sec.~\ref{sec:3}, we formulate the SU(3)TWA for the effective model.
In Sec.~\ref{sec:4}, we study the Gaussian approximation and the DTWA approach for SU(3) phase-space variables. 
In Sec.~\ref{sec:5}, using the SU(3)TWA, we calculate quench dynamics of equal-time single-particle correlation functions for a strongly interacting Bose gas in a 2D optical lattice.
There, we compare some semiclassical results with actual experimental data obtained in Ref.~\cite{Takasu20}.
In Sec.~\ref{sec:6}, we conclude this paper and present outlooks for future studies.

\section{Models} \label{sec:2}

In this paper, we study time evolution of a strongly interacting Bose gas loaded into an optical lattice. 
To describe this system, we consider the Bose-Hubbard Hamiltonian on a certain lattice structure \cite{Fisher89,Jaksch98}
\begin{align}
{\hat H}_{\rm BH} = -J\sum_{\langle j,k \rangle}({\hat a}^{\dagger}_{j}{\hat a}_{k}+{\rm H.c.}) + \frac{U}{2}\sum_{j}{\hat a}^{\dagger}_{j}{\hat a}^{\dagger}_{j}{\hat a}_{j}{\hat a}_{j}, \label{eq: bhm}
\end{align}
where ${\hat a}^{\dagger}_{j}$ and ${\hat a}_{j}$ are the creation and annihilation operators of bosons at site $j$. 
The angular brackets $\langle j, k \rangle$ indicate a nearest-neighbor link on the lattice.
The real parameters $J$ and $U$ denote the hopping amplitude and interaction strength, respectively.
A ratio of the parameters, $U/J$, can be widely controlled by tuning the optical-lattice depth \cite{Takasu20} or utilizing a Feshbach-resonance technique \cite{Braun15}.

In a strongly interacting regime of Eq.~(\ref{eq: bhm}), fluctuations of occupation per site are sufficiently suppressed from the mean filling ${\bar n}$. 
Therefore, only a subset of local Fock states near the mean filling is relevant to strongly interacting dynamics governed by Eq.~(\ref{eq: bhm}). 
If the interaction is sufficiently strong, i.e., $U/({\bar n}J) \gg 1$, one can safely assume that only three Fock states, i.e., $|{\bar n}-1\rangle_{j},|{\bar n}\rangle_{j},|{\bar n}+1\rangle_{j}$ are relevant to time evolution of the interacting bosons.
In a projected Hilbert space spanned by such a local basis, the Bose-Hubbard Hamiltonian~(\ref{eq: bhm}) is approximated as an effective pseudospin-1 model \cite{Huber07,Nagao18}, which is given by
\begin{align}
{\hat H}_{\rm eff} =  &-\frac{{\bar n}J}{2}\sum_{\langle j,k \rangle}(1+\delta\nu_{-}{\hat S}^{z}_{j}){\hat S}^{+}_{j}{\hat S}^{-}_{k}(1+\delta\nu_{-}{\hat S}^{z}_{k})  + {\rm H.c.} \nonumber \\
&+ \frac{U}{2}\sum_{j}({\hat S}^{z}_{j})^2 + \frac{U(2{\bar n}-1)}{2}\sum_{j}{\hat S}^{z}_{j}, \label{eq:effective-model-low}
\end{align}
where $\delta \nu_{-} = \sqrt{1+1/{\bar n}}-1$ and ${\hat S}^{\pm}_{j}={\hat S}^{x}_{j}\pm i{\hat S}^{y}_{j}$. 
The pseudospin operator ${\hat S}^{\mu}_{j}$ ($\mu=x,y,z$) satisfies the SU(2) Lie algebra
\begin{align}
[{\hat S}^{\mu}_{j},{\hat S}^{\nu}_{k}] = i\epsilon_{\mu \nu \gamma}{\hat S}^{\gamma}_{j}\delta_{j,k},\;\;\;\; {\rm for}\;S=1.
\end{align}
The three-leg tensor $\epsilon_{\mu\nu\gamma}$ is the fully antisymmetric structure constant satisfying $\epsilon_{xyz}=-\epsilon_{yxz}=\epsilon_{yzx}=\cdots=1$.
Hereinafter, the repeated greek indices indicate the contraction of tensors.
It should be noticed that if one takes the high-filling limit, i.e., ${\bar n} \gg 1$, the effective model is simplified \cite{Altman02,Nagao16} as
\begin{align}
{\hat H}_{\rm eff}' &= -\frac{J{\bar n}}{2}\sum_{\langle j,k \rangle}({\hat S}^{+}_{j}{\hat S}^{-}_{k}+{\rm H.c.}) + \frac{U}{2}\sum_{j}({\hat S}_{j}^{z})^2 - B\sum_{j}{\hat S}_{j}^{z}, \nonumber
\end{align}
where $B$ can be interpreted as a magnetic field applied along the $z$-axis.
In the previous work \cite{Davidson15}, the SU(3)TWA was applied to this high-filling model defined on a cubic lattice.
However, in order to analyze experimental systems with a setup of ${\bar n}=1$ as realized in Ref.~\cite{Takasu20}, it is required to use Eq.~(\ref{eq:effective-model-low}) rather than the high-filling model.   
In Sec.~\ref{sec:3}, we will explain how one generalizes the SU(3)TWA to Eq.~(\ref{eq:effective-model-low}).

In Sec.~\ref{sec:4}, we present detailed investigations on Monte Carlo integration methods employed for SU(3)TWA simulations.
To examine quantitative validity of our numerical approaches, especially a DTWA approach for SU(3) phase-space variables, we will revisit a fully connected spin-1 model, which is a model studied in Ref.~\cite{Davidson15}.
The Hamiltonian of the fully connected model is given by 
\begin{align}
{\hat H}_{c}=-\frac{J}{2}\sum_{j\neq k}\left[{\hat S}^{x}_{j}{\hat S}^{x}_{k}+{\hat S}^{y}_{j}{\hat S}^{y}_{k}\right]+\frac{U}{2}\sum_{j}({\hat S}^{z}_{j})^{2}. \label{eq: fully-connected model}
\end{align}
The spin-exchange coupling term describes all-to-all connections between distant spin operators.
Hence, each lattice point has a coordination number $z=M-1$. 
As $M$ increases, the valid timescale of the SU(3)TWA for this model becomes longer for a certain $U/(zJ)$ \cite{Davidson15}.
Furthermore, due to a characteristic property described in Appendix~\ref{app:diag}, exact quantum dynamics of this model can be easily simulated by using classical computers even for a considerably large $M$.
Accordingly, the fully connected model is suitable for examining the performance of the sampling methods.
See also Appendix~\ref{app:diag} for details about how to implement exact numerical simulations of this model.

\section{SU(3) truncated-Wigner approximation} \label{sec:3}

The first step for building the SU(3)TWA for spin-1 models is to rewrite their Hamiltonian by means of eight numbers of SU(3) matrices \cite{Davidson15}. 
Let us consider a set of SU(3) generators $\{{\hat X}_{\mu}\}$ ($\mu=1,\cdots,8$) obeying the SU(3) Lie algebra 
\begin{align}
[{\hat X}_{\mu},{\hat X}_{\nu}] &=if_{\mu\nu\gamma}{\hat X}_{\gamma},\;\; \mu,\nu,\gamma = 1,2,\cdots,8.
\end{align}
Here $f_{\mu\nu\gamma}$ is a fully antisymmetric structure constant accompanied by the SU(3) group. 
If we take the Jordan--Schwinger mapping into account, each generator can be written in the bi-linear form of the SU(3) Schwinger bosons \cite{Altman02,Huber07,Nagao18}
\begin{align}
{\hat X}_{\mu} = \sum_{m,n=0}^{2}{\hat b}^{\dagger}_{m}T_{\mu}^{mn}{\hat b}_{n}.
\end{align}
To reproduce the original Hilbert space, the particle number must be preserved per site by a constraint $\sum_{n}{\hat b}^{\dagger}_{n}{\hat b}_{n}=1$. 
The value of $f_{\mu\nu\gamma}$ depends on the detail of $T_{\mu}^{mn}$.
Our choice for $T_{\mu}$ will be shown later in Eq.~(\ref{eq: su3_matrix}), and the corresponding $f_{\mu\nu\gamma}$ will be given by Eq.~(\ref{eq: su3_f}). 
The SU(3) matrices $T_{\mu}$ form a complete set of $3\times 3$ matrices, so that an arbitrary local operator acting on the three-state Hilbert space is expressed as a linear combination of these matrices. 
Using this property, one can linearize local interaction terms in spin-1 models, such as $\frac{U}{2}\sum_{j}({\hat S}^{z}_{j})^2$, in terms of SU(3) matrices.
Specifically for the effective model (\ref{eq:effective-model-low}), if the interaction $U$ is sufficiently large compared to ${\bar n}J$ characterizing the hopping term, then the Hamiltonian is regarded as being almost linear in SU(3) matrices. 
Therefore, the SU(3)TWA for this model is expected to be valid during a long timescale.
Furthermore, if the hopping term is negligible, the SU(3)TWA becomes exact at all times because there exists no truncation error stemming from higher-order derivatives of the time-evolving equation for the Wigner function \cite{Polkovnikov10}.

Let us generalize the SU(3)TWA formalism to the arbitrary filling model (\ref{eq:effective-model-low}). 
First, we express the effective Hamiltonian by means of the local SU(3) generators denoted by ${\hat X}^{(j)}_{\mu}$. 
A key point is that the local interaction term of the SU(2) spin operators is translated into a linear combination of such SU(3) generators as
\begin{align}
\frac{U}{2}({\hat S}^{z}_{j})^2 &\rightarrow \frac{U}{6}\left[ 2{\hat 1}^{(j)}-\sqrt{3}{\hat X}^{(j)}_{8}\right].
\end{align}
Then, we make a Wigner-Weyl transform of the Hamiltonian and obtain a classical Hamiltonian for the SU(3) phase-space variables
\begin{widetext}
\begin{align}
H_{W} = ({\hat H}_{\rm eff})_{W} = &-\frac{{\bar n}J}{4}\delta \nu_{+}^2 \sum_{\langle i,j\rangle}\left[X_{1}^{(i)}X_{1}^{(j)}+X_{2}^{(i)}X_{2}^{(j)}\right]  -\frac{{\bar n}J}{4}\delta \nu_{-}^2 \sum_{\langle i,j\rangle}\left[X_{6}^{(i)}X_{6}^{(j)}+X_{7}^{(i)}X_{7}^{(j)}\right]  \nonumber \\
&+\frac{{\bar n}J}{4}\delta \nu_{+}\delta \nu_{-}\sum_{\langle i,j\rangle}\left[X_{2}^{(i)}X_{7}^{(j)}+X_{7}^{(i)}X_{2}^{(j)}\right] +\frac{{\bar n}J}{4}\delta \nu_{+}\delta \nu_{-}\sum_{\langle i,j\rangle}\left[X_{1}^{(i)}X_{6}^{(j)}+X_{6}^{(i)}X_{1}^{(j)}\right]  \label{eq: classical_h_effective_mod} \\
&-\frac{U}{2\sqrt{3}}\sum_{j}X_{8}^{(j)}+\frac{U(2{\bar n}-1)}{2}\sum_{j}X_{3}^{(j)}, \nonumber
\end{align}
\end{widetext}
where $\delta \nu_{+}=\sqrt{1+1/{\bar n}}+1$. 
The SU(3)TWA states that within a semiclassical approximation the time evolution of the expectation value of an operator ${\hat \Omega}$, i.e., $\langle {\hat \Omega}(t) \rangle$, can be represented in terms of saddle-point trajectories of SU(3) variables, which are governed by $H_{W}$ and weighted with a Wigner quasi-probability distribution function 
\begin{align}
\langle {\hat \Omega}(t) \rangle \approx \int d^{8}\bm{X}_0 W(\bm{X}_0)\Omega_{W}[\bm{X}_{\rm cl}(t)],
\end{align}
where $d^{8}\bm{X}_0 = \prod_{j,\mu}dX^{(j)}_{0,\mu}$ is the integration measure and $\Omega_{W}$ is a Weyl symbol of ${\hat \Omega}$.
The classical trajectory $\bm{X}_{\rm cl}(t)$ obeys Hamilton's equation associated with the SU(3) Lie algebra
\begin{align}
\hbar {\dot X}^{(j)}_{\mu} = f_{\mu \nu \gamma}\frac{\partial H_{W}}{\partial X^{(j)}_{\nu}}X^{(j)}_{\gamma}. \label{eq: su3_eom_multisite}
\end{align}
This equation of motion is integrated under an initial condition $X^{(j)}_{\mu}(t=0)=X^{(j)}_{0,\mu}$. 
The $c$ number $X^{(j)}_{0,\mu}$ is distributed according to $W({\bm X}_0)$.
The width of the Wigner function gives quantum-fluctuation corrections to saddle-point or mean-field results, which formally correspond to the time-dependent Gutzwiller approximation with a single-site cluster consisting of three levels. 

If we take the high-filling limit for the classical Hamiltonian (\ref{eq: classical_h_effective_mod}), all the terms involving $X^{(j)}_{6}$ and $X^{(j)}_{7}$ disappear. 
Therefore, these additional variables are responsible for different consequences between the high and low filling descriptions. 
It should be noted that a constant term has been eliminated from Eq.~(\ref{eq: classical_h_effective_mod}) because it does not affect Eq.~(\ref{eq: su3_eom_multisite}).
The above formalism will be used in Sec.~\ref{sec:5} to analyze the experimental setup in Ref.~\cite{Takasu20}.

In this paper, we will utilize the following representation for the SU(3) matrices, according to the notations by Davidson and Polkovnikov \cite{Davidson15}:
\begin{align}
T_1&=
\begin{bmatrix}
0                            &\frac{1}{\sqrt{2}}    &0 \\
\frac{1}{\sqrt{2}}     &0                           &\frac{1}{\sqrt{2}} \\
0                            &\frac{1}{\sqrt{2}}    &0
\end{bmatrix}
,\;\;
T_2=
\begin{bmatrix}
0                                    &-\frac{i}{\sqrt{2}}    & 0 \\
\frac{i}{\sqrt{2}}     & 0                                   &-\frac{i}{\sqrt{2}} \\
0                                    & \frac{i}{\sqrt{2}}    & 0
\end{bmatrix} 
, \nonumber \\
T_3&=
\begin{bmatrix}
1     &0     & 0 \\
0     &0     & 0 \\
0     &0     &-1
\end{bmatrix}
,\;\;
T_4=
\begin{bmatrix}
0     &0     & 1 \\
0     &0     & 0 \\
1     &0     & 0
\end{bmatrix}
,  \label{eq: su3_matrix} \\
T_5&=
\begin{bmatrix}
0            &0     & - i \\
0            &0     &          0 \\
i     &0     &          0
\end{bmatrix}
,\;\; 
T_6=
\begin{bmatrix}
0                           &-\frac{1}{\sqrt{2}}   & 0 \\
-\frac{1}{\sqrt{2}}  &0                           &  \frac{1}{\sqrt{2}}  \\
0                            &\frac{1}{\sqrt{2}}    &          0
\end{bmatrix}
,\nonumber \\
T_7&=
\begin{bmatrix}
0                                  &\frac{i}{\sqrt{2}}  & 0 \\
-\frac{i}{\sqrt{2}}  &0                                 &  -\frac{i}{\sqrt{2}}  \\
0                                  &\frac{i}{\sqrt{2}}  &          0
\end{bmatrix}
,\;\;
T_8=
\begin{bmatrix}
-\frac{1}{\sqrt{3}}    & 0                           &                         0 \\
0                             & \frac{2}{\sqrt{3}}   &                         0 \\
0                             & 0                          &   -\frac{1}{\sqrt{3}}
\end{bmatrix}
.\nonumber
\end{align}
These matrices are normalized as
\begin{align}
{\rm Tr}\left[ T_{\mu}T_{\nu} \right] = 2\delta_{\mu,\nu}.
\end{align}
It is confirmed that, in this specific representation, non-zero values of $f_{\mu \nu \gamma}$ are given by
\begin{align}
f_{123} = f_{147} = f_{165} = f_{246} = f_{257} = f_{367} = 1, \nonumber \\
f_{178} = f_{286} = \sqrt{3},\;\;\;\;\;\;\;\;\;\;\;\;\;\;\;\;\;\;\;\;\;\; \label{eq: su3_f} \\
f_{345} = 2. \;\;\;\;\;\;\;\;\;\;\;\;\;\;\;\;\;\;\;\;\;\;\;\;\;\;\;\;\;   \nonumber
\end{align}
Of course, this is not the unique choice. Instead of this representation, one can also use the Gell-Mann matrices, which are more familiar in high-energy physics \cite{Georgi18}.

\section{Monte Carlo integrations} \label{sec:4}

In this section, we study Monte Carlo integration methods for evaluating the phase-space integration of the initial Wigner function. 
In Ref.~\cite{Davidson15}, an approximate Gaussian-Wigner function has been used to perform numerical simulations.
This Gaussian approximation is a simple and efficient prescription for resolving a kind of minus-sign problem in TWA simulations, which means that the exact Wigner function defined by means of the Schwinger-boson coherent states typically takes negative values.
In Sec.~\ref{sec:4}, to simulate the experimental setup, we will indeed employ the Gaussian approach.

As an alternative sampling scheme that allows us to avoid the appearance of negative-valued Wigner function, we also use a DTWA approach \cite{Schachenmayer15}. 
This approach is formulated on the basis of the discrete-Wigner representation of a finite Hilbert space quantum system. 
The concept of the discrete-Wigner representation has been invented by Wootters in Ref.~\cite{Wootters87}. 
In this section, by extending the previous DTWA method for SU(2) spin systems \cite{Schachenmayer15}, we develop a DTWA approach suited for the SU(3)TWA.
To this end, we will introduce phase-point operators for the SU(3) generators, each of which is represented as a $3\times 3$ matrix. 

\subsection{Gaussian approximation}

In the Gaussian approximation for exact Wigner functions, an appropriate Gauss distribution is used to approximately express initial density matrices within a class of positive-definite functions \cite{Davidson15}. 
To be specific, let us consider a fully polarized state along the $x$ axis, i.e., ${\hat \rho}_{1} = |S_{x}=1\rangle \langle S_{x}=1|$. 
Its matrix form is given by
\begin{align}
\rho_{1} &= 
\begin{bmatrix}
\frac{1}{4} & \frac{1}{2\sqrt{2}} & \frac{1}{4} \\
\frac{1}{2\sqrt{2}} & \frac{1}{2} & \frac{1}{2\sqrt{2}} \\
\frac{1}{4} & \frac{1}{2\sqrt{2}} & \frac{1}{4} 
\end{bmatrix}.\label{eq:initial_s_x=1}
\end{align} 
To obtain the corresponding Gaussian-Wigner function, we make the following ansatz with free parameters $R=(R_{\mu\nu})$, $\bm{m}=(m_\mu)$, and $\bm{\sigma}=(\sigma_\mu)$:
\begin{align}
P(\{X_{\mu}\}) = \prod_{\mu=1}^{8}\frac{1}{\sqrt{2\pi} \sigma_{\mu}}e^{-\frac{1}{2\sigma^2_{\mu}}\left(R_{\mu\nu}X_{\nu}-m_{\mu}\right)^2}. \label{eq: gauss}
\end{align}
This distribution defines the first and second order moments of the SU(3) phase-space variables
\begin{align}
\overline{X_{\mu}} &= \int d^8 X P(\{X_{\mu}\}) X_{\mu}, \\
\overline{X_{\mu}X_{\nu}} &= \int d^8 X P(\{X_{\mu}\}) X_{\mu}X_{\nu}.
\end{align}
The free parameters are determined such that the Gaussian-Wigner function exactly reproduces the first- and second-moments of the density matrix (\ref{eq:initial_s_x=1}), i.e.,
\begin{align}
\overline{X_{\mu}} &\equiv \langle {\hat X}_{\mu} \rangle, \\
\overline{X_{\mu}X_{\nu}} &\equiv \frac{1}{2}\langle {\hat X}_{\mu}{\hat X}_{\nu} + {\hat X}_{\nu}{\hat X}_{\mu} \rangle.
\end{align}
The angular brackets in the right-hand side mean the quantum-mechanical average with ${\hat \rho}_{1}$. 
To determine $R$ in practice, we diagonalize an $8 \times 8$ matrix corresponding to a connected and symmetrized correlation function with respect to the density matrix
\begin{align}
C_{\mu\nu} = \frac{1}{2}\langle {\hat X}_{\mu}{\hat X}_{\nu} + {\hat X}_{\nu}{\hat X}_{\mu} \rangle - \langle {\hat X}_{\mu} \rangle\langle {\hat X}_{\nu} \rangle. 
\end{align}
The eight-dimensional matrix $R$ is constructed from the eigenvectors, which are obtained when $C_{\mu\nu}$ is diagonalized. 
Each eigenvalue gives the squared covariance $\sigma^2_{\mu}$. 
The mean value $m_{\mu}$ is the rotation of the vector $(\langle {\hat X}_{\mu} \rangle)$, i.e., $R_{\mu\nu}\langle {\hat X}_{\nu} \rangle = m_{\mu}$. 
The direct calculation leads to the following result:
\begin{align}
R &= 
\begin{bmatrix}
0 & 0 & 0 & -\frac{1}{2} & 0 & 0 & 0 & \frac{\sqrt{3}}{2} \\
0 & 0 & 0 & 0 & 0 & 0 & 1 & 0 \\
0 & 0 & -\frac{1}{\sqrt{2}} & 0 & 0 & \frac{1}{\sqrt{2}} & 0 & 0 \\
0 & \frac{1}{\sqrt{2}} & 0 & 0 & \frac{1}{\sqrt{2}} & 0 & 0 & 0 \\
0 & 0 & 0 & \frac{\sqrt{3}}{2} & 0 & 0 & 0 & \frac{1}{2} \\
0 & 0 & \frac{1}{\sqrt{2}} & 0 & 0 & \frac{1}{\sqrt{2}} & 0 & 0 \\
0 & -\frac{1}{\sqrt{2}} & 0 & 0 & \frac{1}{\sqrt{2}} & 0 & 0 & 0 \\
1 & 0 & 0 & 0 & 0 & 0 & 0 & 0
\end{bmatrix}, \label{eq:gauss_x_R} \\
\bm{m} &= 
\begin{bmatrix}
0 & 0 & 0 & 0 & \frac{1}{\sqrt{3}} & 0 & 0 & 1
\end{bmatrix}^{T}, \label{eq:gauss_x_m} \\
\bm{\sigma} &= 
\begin{bmatrix}
1 & 1 & 1 & 1 & 0 & 0 & 0 & 0
\end{bmatrix}^{T}.  \label{eq:gauss_x_sigma}
\end{align}
With these parameters, the Gauss distribution (\ref{eq: gauss}) randomly generates the phase-space variables reproducing the exact low-order moments of the state in Eq.~(\ref{eq:initial_s_x=1}).

In the projected Hilbert space for the effective pseudospin-1 models, the deep Mott-insulator state, which is approximately realized in a sufficiently deep optical lattice, is expressed as a direct product state of ${\hat \rho}_{2} = |S_{z}=0\rangle\langle S_{z}=0|$. 
The matrix form of ${\hat \rho}_{2}$ is given by
\begin{align}
\rho_{2} &= 
\begin{bmatrix}
0 & 0 & 0 \\
0 & 1 & 0 \\
0 & 0 & 0 
\end{bmatrix}. \label{eq: density_matrix_sz_0}
\end{align}
The corresponding parameters of the Gauss distribution function are calculated as
\begin{align}
R &= 
\begin{bmatrix}
0 & 0 & 0 & 0 & 0 & 0 & 1 & 0 \\
0 & 0 & 0 & 0 & 0 & 1 & 0 & 0 \\
0 & 1 & 0 & 0 & 0 & 0 & 0 & 0 \\
1 & 0 & 0 & 0 & 0 & 0 & 0 & 0 \\
0 & 0 & 0 & 0 & 0 & 0 & 0 & 1 \\
0 & 0 & 0 & 0 & 1 & 0 & 0 & 0 \\
0 & 0 & 0 & 1 & 0 & 0 & 0 & 0 \\
0 & 0 & 1 & 0 & 0 & 0 & 0 & 0
\end{bmatrix}, \\
\bm{m} &= 
\begin{bmatrix}
0 & 0 & 0 & 0 & \frac{2}{\sqrt{3}} & 0 & 0 & 0
\end{bmatrix}^{T},\\
\bm{\sigma} &= 
\begin{bmatrix}
1 & 1 & 1 & 1 & 0 & 0 & 0 & 0
\end{bmatrix}^{T}.
\end{align}

\subsection{SU(3) discrete-Wigner representation} \label{sec: discrete wigner}

Let us consider a discrete-Wigner representation for a finite-level system, whose Hilbert space is spanned by three basis vectors $\{|0\rangle,|1\rangle,|2\rangle\}$. 
The key building blocks for this representation are the so-called phase-point operators ${\hat A}_{\alpha}$, which are $3\times 3$ matrices acting on the Hilbert space. 
The integer index $\alpha=(a_1,a_2)$ ($a_{1},a_{2}=0,1,2$) expresses a point in the discrete phase space $\Gamma$, which now contains nine points. 
The phase-point operators are also called the Stratonovich-Weyl kernels \cite{Brif99}.

\begin{figure*}
\includegraphics[width=160mm]{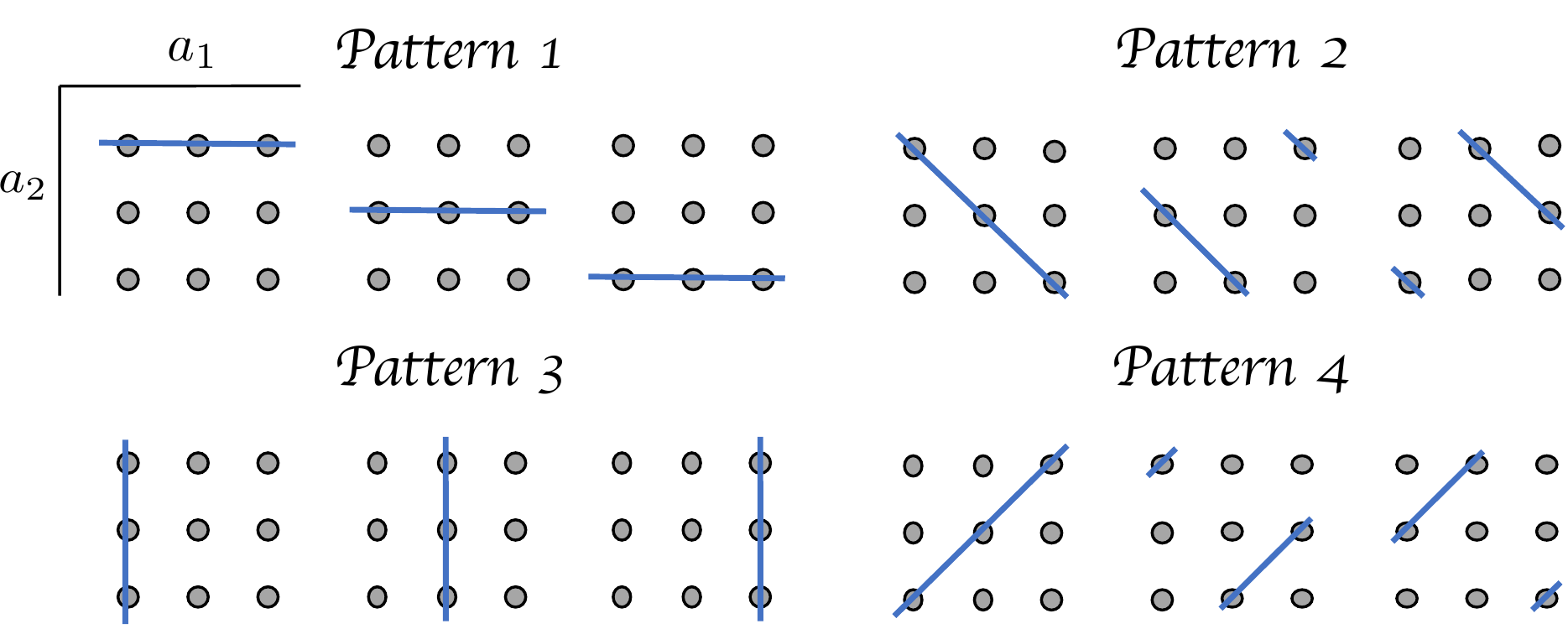}
\vspace{0mm}
\caption{Possible parallel lines in $\Gamma$. 
The boundaries of each $3\times 3$ square are periodic. 
The vertical and horizontal axes are $a_2$ and $a_1$, respectively.
See also Ref.~\cite{Wootters87}.
}
\label{figure01}
\end{figure*}

The phase-point operators are important because they define a Wigner-Weyl transform of quantum-mechanical operators. 
In the discrete-Wigner representation, the Weyl symbol of an operator ${\hat \Omega}$ is defined as its projection onto a point $\alpha \in \Gamma$:
\begin{align}
\Omega_{\alpha} = {\rm Tr}[{\hat A}_{\alpha}{\hat \Omega}]. \label{eq:omega_dw}
\end{align}
Specifically, such a projection of a given density matrix ${\hat \rho}$ leads to the discrete-Wigner function
\begin{align}
w_{\alpha} = \frac{1}{3}{\rm Tr}[{\hat \rho}{\hat A}_{\alpha}]. \label{eq:wigner_dw}
\end{align}
The pre-factor $1/3$ is needed to ensure the unity normalization of the Wigner function $\sum_{\alpha \in \Gamma}w_{\alpha}=1$, see also below. 
By analogy with continuous cases, where the coordinate and momentum operators $({\hat x},{\hat p})$ define a {\it continuous} phase-point operator, the {\it discrete} phase-point operator should have the following properties \cite{Wootters87}:
\begin{enumerate}
\item{{\it Hermiticity}: ${\hat A}_{\alpha}^{\dagger} = {\hat A}_{\alpha}$ for any $\alpha \in \Gamma$. Then, the phase space functions are real as long as the corresponding operators are Hermitian.}
\item{{\it Normalization}: ${\rm Tr}[{\hat A}_{\alpha}] = 1$ for any $\alpha \in \Gamma$. This means that the Weyl symbol of the unit operator ${\hat 1}$ is set to unity: $({\hat 1})^{W}_{\alpha} = 1$.}
\item{{\it Orthogonality with respect to the trace inner product}: ${\rm Tr}[{\hat A}_{\alpha}{\hat A}_{\alpha'}] = 3 \delta_{a_1,a'_1}\delta_{a_2,a'_2}$ for $\alpha,\alpha' \in \Gamma$. Here $\delta_{a,a'}$ is Kronecker's delta.}
\item{{\it Projection operators on parallel lines}: For the three-state case, there are four different patterns of drawing three parallel lines on $\Gamma$ (Fig.~\ref{figure01}). 
For each line $l$ involving three points, one can make a projection operator ${\hat P}_{l}=3^{-1}\sum_{\alpha \in l}{\hat A}_{\alpha}$. Then, ${\hat P}_{l_1}{\hat P}_{l_2}=0$ if $l_1 \parallel l_2$ and $l_1 \neq l_2$. The sum of the projectors is equal to unity: $\sum_{l}{\hat P}_{l} = \sum_{\alpha \in\Gamma}{\hat A}_{\alpha}=1$.}
\end{enumerate}
Such discrete phase-point operators can also be made for general cases where the Hilbert space is in ${\cal N}$ dimensions (${\cal N} \geq 2$ is a primal number) \cite{Wootters87}. 
Furthermore, it is possible to construct a {\it discrete} number-phase representation for Bose systems, whose Hilbert space is spanned by generators of the Heisenberg-Weyl group, and it provides a DTWA-like semiclassical approximation for their quantum dynamics if the allowed occupancy of particles is sufficiently large \cite{Hush10}.

As an inverse transformation of Eqs.~(\ref{eq:omega_dw}) and (\ref{eq:wigner_dw}), the operators ${\hat \Omega}$ and ${\hat \rho}$ are linearly expanded in ${\hat A}_{\alpha}$ such that
\begin{align}
{\hat \Omega} = \frac{1}{3}\sum_{\alpha \in \Gamma}\Omega_{\alpha}{\hat A}_{\alpha},\;\; {\hat \rho} = \sum_{\alpha \in \Gamma}w_{\alpha}{\hat A}_{\alpha}.
\end{align}
Then, the expectation value of ${\hat \Omega}$ for ${\hat \rho}$ reads as
\begin{align}
\langle {\hat \Omega} \rangle = {\rm Tr}[{\hat \rho}{\hat \Omega}] = \sum_{\alpha \in \Gamma}w_{\alpha}\Omega_{\alpha}.
\end{align}
The summation in the last expression is taken over the whole $\Gamma$. In the second equality, we have used the trace orthogonality of ${\hat A}_{\alpha}$.

The concrete forms of $\Omega_{\alpha}$ and $w_{\alpha}$ are specified after one determines ${\hat A}_{\alpha}$ for all $\alpha=(a_1,a_2)$ such that they satisfy the required conditions as presented above. 
If we adopt Wootters's representation of the phase-point operators \cite{Wootters87}, we have
\begin{align}
A^{(0)}_{\alpha} = 
\begin{bmatrix}
\delta_{a_1,0}&\delta_{a_1,2}e^{-i\frac{2\pi a_2}{3}}&\delta_{a_1,1}e^{-i\frac{4\pi a_2}{3}} \\
\delta_{a_1,2}e^{i\frac{2\pi a_2}{3}}&\delta_{a_1,1}&\delta_{a_1,0}e^{-i\frac{2\pi a_2}{3}} \\
\delta_{a_1,1}e^{i\frac{4\pi a_2}{3}}&\delta_{a_1,0}e^{i\frac{2\pi a_2}{3}}&\delta_{a_1,2} 
\end{bmatrix}.
\end{align}
It is convenient to expand ${\hat A}_{\alpha}$ in the generators of the SU(3) Lie algebra, i.e.,
\begin{align}
{\hat A}_{\alpha} = \frac{1}{3}\left( {\hat 1} + \frac{3}{2}x_{\mu}(\alpha){\hat X}_{\mu} \right).
\end{align}
Its projection coefficient $x_{\mu}(\alpha) = {\rm Tr}[{\hat A}_{\alpha}{\hat X}_{\mu}]$ is the discrete Weyl symbol of ${\hat X}_{\mu}$. 
After direct calculations, we obtain the following discrete phase-space variables for $A^{(0)}_{\alpha}$:
\begin{align}
x_{1}(\alpha)&= \sqrt{2}\delta_{a_1,0}{\rm cos}\frac{2\pi a_2}{3} + \sqrt{2}\delta_{a_1,2}{\rm cos}\frac{2\pi a_2}{3}, \nonumber \\
x_{2}(\alpha)&= \sqrt{2}\delta_{a_1,0}{\rm sin}\frac{2\pi a_2}{3} + \sqrt{2}\delta_{a_1,2}{\rm sin}\frac{2\pi a_2}{3}, \nonumber \\
x_{3}(\alpha)&=\delta_{a_1,0} - \delta_{a_1,2}, \nonumber \\
x_{4}(\alpha)&=2\delta_{a_1,1}{\rm cos}\frac{4\pi a_2}{3}, \nonumber \\
x_{5}(\alpha)&=2\delta_{a_1,1}{\rm sin}\frac{4\pi a_2}{3},  \\
x_{6}(\alpha)&=-\sqrt{2}\delta_{a_1,2}{\rm cos}\frac{2\pi a_2}{3} + \sqrt{2}\delta_{a_1,0}{\rm cos}\frac{2\pi a_2}{3}, \nonumber \\
x_{7}(\alpha)&=-\sqrt{2}\delta_{a_1,2}{\rm sin}\frac{2\pi a_2}{3} + \sqrt{2}\delta_{a_1,0}{\rm sin}\frac{2\pi a_2}{3}, \nonumber \\
x_{8}(\alpha)&=-\frac{1}{\sqrt{3}}(\delta_{a_1,0}+\delta_{a_1,2})+\frac{2}{\sqrt{3}}\delta_{a_1,1}. \nonumber
\end{align}
Notice that different values of $\alpha=(a_1,a_2)$ correspond to different configurations of the SU(3) phase-space variables. 
For example, if we write $\bm{x}(\alpha)=[x_1(\alpha),\cdots,x_8(\alpha)]$ as a combined eight-dimensional vector on each phase point, $\alpha=(0,1)\text{, }(1,2)\text{, and }(2,0)$ correspond to the following configurations, respectively:
\begin{align}
\bm{x}(0,1) &=
\begin{bmatrix}
\frac{-1}{\sqrt{2}} & \sqrt{\frac{3}{2}} & 1 & 0 & 0 & \frac{-1}{\sqrt{2}} & \sqrt{\frac{3}{2}} & \frac{-1}{\sqrt{3}} 
\end{bmatrix}, \nonumber \\
\bm{x}(1,2) &= 
\begin{bmatrix}
0 & 0 & 0 & -1 & \sqrt{3} & 0 & 0 & \frac{2}{\sqrt{3}}
\end{bmatrix}, \nonumber \\
\bm{x}(2,0) &=
\begin{bmatrix}
\sqrt{2} & 0 & -1 & 0 & 0 & -\sqrt{2} & 0 & \frac{-1}{\sqrt{3}}
\end{bmatrix}. \nonumber 
\end{align}
Two classical spins $\bm{x}(\alpha)$ and $\bm{x}(\alpha')$ at different points $\alpha \neq \alpha'$ are not orthogonal to each other. 
Indeed, these have a finite inner product even for $\alpha \neq \alpha'$
\begin{align}
\bm{x}(\alpha)\cdot \bm{x}(\alpha') 
&= \frac{16}{3}\delta_{\alpha,\alpha'} - \frac{2}{3}(1-\delta_{\alpha,\alpha'}) \nonumber \\
&= 6\delta_{\alpha,\alpha'} - \frac{2}{3}.
\end{align}
In the DTWA simulation, such discretized spins are randomly distributed according to $w_{\alpha}$ and give a set of initial conditions for the classical trajectories. 
The discussions of the DTWA for the SU(3) systems will be presented in Sec.~\ref{sec:dtwa}. 

To clarify the sampling weight of DTWA simulations, which will be used in the following sections, let us calculate the discrete Wigner function for the Mott insulator state [Eq.~(\ref{eq: density_matrix_sz_0})] by using $A^{(0)}_{\alpha}$. 
It results in a positive-definite distribution function 
\begin{align}
w^{(0)}_{\alpha} = \frac{1}{3}{\rm Tr}[\rho_{2}A^{(0)}_{\alpha}] = \frac{1}{3}\delta_{a_1,1}. \label{eq:dwigner_mott}
\end{align}
This result means that in the Mott-insulator state three configurations at $\alpha=(1,0),(1,1),(1,2)$ are realized with equal probability $\frac{1}{3}$ while other ones have the zero probability. 
Therefore, we can directly evaluate the average with the Wigner function in numerics without further approximation of the distribution function. 
However, the positivity of Eq.~(\ref{eq:dwigner_mott}) is not a general property. 
For example, the $x$-polarized state in Eq.~(\ref{eq:initial_s_x=1}) yields oscillatory terms in the distribution 
\begin{align}
{w'}^{(0)}_{\alpha} &= \frac{1}{3}{\rm Tr}[\rho_{1}A^{(0)}_{\alpha}] \nonumber \\
&= \frac{\delta_{a_1,1}}{6}\left[ 1 + {\rm cos}\frac{4\pi a_2}{3} \right] \nonumber \\
&\;\;\;\;\;\;\;\;\; + \frac{\delta_{a_1,0}+\delta_{a_1,2}}{12}\left[ 1 + 2\sqrt{2}{\rm cos}\frac{2\pi a_2}{3} \right].
\end{align}
While the first term with $\delta_{a_1,1}$ is always positive, the second term with $\delta_{a_1,0}$ and $\delta_{a_1,2}$ takes negative values due to the oscillating contributions. 

As mentioned in previous works \cite{Wootters87,Pucci16}, the definition of the phase-point operators is not unique.
In general, there exists a non-singular (or regular) transformation, ${\hat A}_{\alpha} \rightarrow {\hat S}^{-1} {\hat A}_{\alpha} {\hat S}$, which retains the required properties of the phase-point operators \cite{Wootters87}. 
This type of ambiguity will be utilized in Appendix \ref{app:ppo} to construct a reasonable set of phase-point operators for given density matrices.

\subsection{SU(3)DTWA} \label{sec:dtwa}

Here we formulate the DTWA for the SU(3) phase-space variables. 
Throughout this paper, we refer to this approach as the SU(3)DTWA. 

Let us consider real-time dynamics of a many-body spin-1 system described by a Hamiltonian ${\hat H}$. 
The initial density matrix ${\hat \rho}_0 = {\hat \rho}(t=0)$ can be expressed as an expansion in a tensor product of local phase-point operators 
\begin{align}
{\hat \rho}_0 = \sum_{\bm{\alpha}\in \Gamma^{M}} w_{\bm{\alpha}} {\hat A}_{\alpha_1}\otimes \cdots \otimes {\hat A}_{\alpha_M},
\end{align}
where $w_{\bm{\alpha}} \equiv w_{\alpha_1,\cdots,\alpha_{M}}$ is a many-body discrete-Wigner function defined in the $M$-body phase space $\Gamma^{M} \equiv \Gamma_{1} \otimes \cdots \otimes \Gamma_{M}$. 
Note that $M$ typically represents a total number of sites for lattice systems.
Each local operator ${\hat A}_{\alpha_j}$ acts on the site $j$. Such an expansion is expected to exist for any states because a set of ${\hat A}_{\alpha_j}$ forms a local operator basis. 
An operator ${\hat \Omega}$ that we are interested in has also an expansion given by
\begin{align}
{\hat \Omega} = \frac{1}{3^{M}}\sum_{\bm{\alpha}\in\Gamma^{M}}\Omega_{\bm{\alpha}}{\hat A}_{\alpha_1}\otimes \cdots \otimes {\hat A}_{\alpha_M}.
\end{align}
Then, the expectation value of ${\hat \Omega} $ at time $t>0$, i.e., $\langle {\hat \Omega}(t)\rangle = {\rm Tr}[{\hat \Omega}{\hat U}(t){\hat \rho}_0{\hat U}^{\dagger}(t)]$ reads as
\begin{align}
\langle {\hat \Omega}(t) \rangle &= \frac{1}{3^{M}}\sum_{\bm{\alpha} \in \Gamma^{M}} \sum_{\bm{\beta} \in \Gamma^{M}} \Omega_{\bm{\beta}} {\cal U}_{W}(\bm{\beta},\bm{\alpha};t) w_{\bm{\alpha}}.
\end{align}
The propagation function ${\cal U}_{W}(\bm{\beta},\bm{\alpha};t)$ connecting two Weyl symbols $\Omega_{\bm{\beta}}$ and $w_{\bm{\alpha}}$ is defined by
\begin{align}
{\cal U}_{W}(\bm{\beta},\bm{\alpha};t) = {\rm Tr}[ {\hat {\mathscr A}}_{\bm{\beta}} {\hat U}(t) {\hat {\mathscr A}}_{\bm{\alpha}} {\hat U}^{\dagger}(t)],
\end{align}
where ${\hat {\mathscr A}}_{\bm{\alpha}} = \bigotimes_{j=1}^{M}{\hat A}_{\alpha_j}$ and ${\hat U}(t)=e^{-\frac{i}{\hbar}{\hat H}t}$ is the unitary time-evolution operator. 
This propagator contains complete information of quantum many-body dynamics governed by ${\hat H}$.
However, the unitary transformation given by ${\hat U}(t) {\hat {\mathscr A}}_{\bm{\alpha}} {\hat U}^{\dagger}(t)$ changes the tensor product into complicated operator strings in the Hilbert space, so that the exact evaluation of ${\cal U}_{W}(\bm{\beta},\bm{\alpha};t)$ is generally impossible.

The TWA for quantum dynamics is nothing else but an appropriate semiclassical approximation for the phase-space propagator ${\cal U}_{W}(\bm{\beta},\bm{\alpha};t)$ \cite{Berg09}. 
In the treatment discussed in Ref.~\cite{Schachenmayer15}, one makes the following direct-product ansatz for the many-body phase-point operators at time $t>0$:
\begin{align}
{\hat U}(t) {\hat {\mathscr A}}_{\bm{\alpha}} {\hat U}^{\dagger}(t) \approx {\hat A}_{1}[\bm{x}^{(1)}(t)] \otimes \cdots \otimes {\hat A}_{M}[\bm{x}^{(M)}(t)], \nonumber
\end{align}
where 
\begin{align}
{\hat A}_{j}[\bm{x}^{(j)}(t)] = \frac{1}{3}\left[ {\hat 1}^{(j)} + \frac{3}{2}x^{(j)}_{\mu}(t){\hat X}^{(j)}_{\mu} \right].
\end{align}
The time dependence of $x^{(j)}_{\mu}(t)$ is determined by a set of classical equations of motion with initial conditions $x^{(j)}_{\mu}(t=0)={\rm Tr}[{\hat {\mathscr A}}_{\bm \alpha}{\hat X}^{(j)}_{\mu}]\equiv x^{(j)}_{0,\mu}(\alpha_{j})$, which has the form
\begin{align}
\hbar \frac{\partial x_{\mu}^{(j)}(t)}{\partial t} = f_{\mu\nu\rho}\frac{\partial {\cal H}_{W}}{\partial x^{(j)}_{\nu}} x^{(j)}_{\rho}. \label{eq: eom_dtwa}
\end{align}
The classical Hamiltonian ${\cal H}_{W}$ can be derived by replacing ${\hat X}^{(j)}_{\mu}$ of ${\hat H}$ with corresponding phase-space variables. 
At least formally, ${\cal H}_{W}$ coincides with the {\it continuous} Weyl symbol of ${\hat H}$ (see also Sec.~\ref{sec:3}), so that Eq.~(\ref{eq: eom_dtwa}) is equivalent to Eq.~(\ref{eq: su3_eom_multisite}) in the continuous SU(3)TWA. 
Thus, the propagator is approximated as a direct product of trace inner products
\begin{align}
{\cal U}_{W}(\bm{\beta}, \bm{\alpha};t) \approx \prod_{j=1}^{M}{\rm Tr}\left\{{\hat A}_{\beta_{j}}{\hat A}_{j}[\bm{x}^{(j)}(t;{\bm \alpha})]\right\}, \label{eq: approximate_propagator}
\end{align}
where $x^{(j)}_{\mu}(t=0)={\rm Tr}[{\hat {\mathscr A}}_{\bm \alpha}{\hat X}^{(j)}_{\mu}]$.
Each local part simply results in an inner product of two vectors, i.e.,
\begin{align}
{\rm Tr}\left\{{\hat A}_{\beta_{j}}{\hat A}_{j}[\bm{x}^{(j)}(t;{\bm \alpha})]\right\} = \frac{1}{3}+\frac{1}{2} \bm{x}^{(j)}_{0}(\beta_{j}) \cdot \bm{x}^{(j)}(t;{\bm \alpha}). \nonumber
\end{align}

Inserting Eq.~(\ref{eq: approximate_propagator}), we finally arrive at the SU(3)DTWA representation of $\langle {\hat \Omega}(t) \rangle$:
\begin{align}
\langle {\hat \Omega}(t) \rangle &\approx \frac{1}{3^{M}}\sum_{ \bm{\alpha} \in \Gamma^{M}} \sum_{ \bm{\beta} \in \Gamma^{M}} w_{\bm{\alpha}}\Omega_{\bm{\beta}} \nonumber \\
&\;\;\times \prod_{j=1}^{M}\left[\frac{1}{3}+\frac{1}{2} \bm{x}^{(j)}_{0}(\beta_{j}) \cdot \bm{x}^{(j)}(t;{\bm \alpha}) \right].
\end{align}
If we put ${\hat \Omega} = {\hat X}^{(j)}_{\mu}$ or ${\hat \Omega} = {\hat X}^{(j)}_{\mu}{\hat X}^{(k)}_{\nu}$ ($j \neq k$) and perform the summation over $\bm{\beta} \in \Gamma^{M}$, we have the formulas
\begin{align}
\langle {\hat X}^{(j)}_{\mu}(t) \rangle &\approx \sum_{\bm{\alpha} \in \Gamma^{M}} w_{\bm{\alpha}} x^{(j)}_{\mu}(t;{\bm \alpha}), \nonumber \\
\langle {\hat X}^{(j)}_{\mu}(t){\hat X}^{(k)}_{\nu}(t) \rangle &\approx \sum_{\bm{\alpha} \in \Gamma^{M}} w_{\bm{\alpha}} x^{(j)}_{\mu}(t;{\bm \alpha})x^{(k)}_{\nu}(t;{\bm \alpha}). \nonumber
\end{align}
In typical cases, initial density matrices are factorized with respect to the single-body index $j$. 
Then, the discrete-Wigner function reads as
\begin{align}
w_{\bm{\alpha}} = \prod_{j=1}^{M}w_{\alpha_j}.
\end{align}
Therefore, we obtain
\begin{align}
\langle {\hat X}^{(j)}_{\mu}(t) \rangle &\approx \prod_{l=1}^{M}\sum_{ \alpha_{l} \in \Gamma_{l}} w_{\alpha_{l}} x^{(j)}_{\mu}(t;{\bm \alpha}), \nonumber \\
\langle {\hat X}^{(j)}_{\mu}(t){\hat X}^{(k)}_{\nu}(t) \rangle &\approx \prod_{l=1}^{M}\sum_{ \alpha_{l} \in \Gamma_{l}} w_{\alpha_{l}} x^{(j)}_{\mu}(t;{\bm \alpha})x^{(k)}_{\nu}(t;{\bm \alpha}). \nonumber
\end{align}

As learned from these expressions, the only difference of the SU(3)DTWA from the standard SU(3)TWA comes from their probability distributions for the phase-space variables. 
In other words, the classical dynamics in the SU(3)DTWA still happen in the continuous phase space.
Compared to the Gaussian approximation, the DTWA method features a numerical advantage that it allows to sample spin configurations with positive probabilities for typical product states, which give rise to negative probabilities in the exact continuous representation~\cite{Schachenmayer15,kunimi2021performance}. 
In the literature such as Ref.~\cite{Schachenmayer15}, examples are presented, demonstrating that the DTWA improves revival properties of the quantum dynamics, which the Gaussian approximation fails to capture.  
The direct comparison between the two methods will be presented in Sec.~\ref{sec:small_size}.

We mention that our description, which explicitly uses the phase-point operators and, therefore, explicitly defines a discrete-Wigner function for a density matrix, is distinct from a similar discrete-sampling approach for general SU(${\cal N}$) systems developed in Ref.~\cite{Zhu19}.
The latter approach has not introduced any phase-point operators explicitly, but instead has utilized a quantum-tomography-like methodology to define probability distributions for each phase-space variable. 
This state-of-the-art sampling technique, which is also called the generalized DTWA (GDTWA)~\cite{Zhu19}, has been already applied to actual experimental setups of large-spin systems such as $^{52}{\rm Cr}$ gases~\cite{lepoutre2019out} and $^{167}{\rm Er}$ gases~\cite{patscheider2020controlling}, and the performance has been evaluated against the experimental data.
In Sec.~\ref{sec:5}, we compare this sampling scheme to our schemes, specifically for the 2D Bose-Hubbard model with a small size.

To implement the tomography technique for the SU(3) TWA, we decompose each SU(3) matrix $T_{\mu}$ in its diagonalized basis, i.e., $T_{\mu}=\sum_{s=1}^{3}\lambda^{(s)}_{\mu}|\phi_{\mu}^{(s)}\rangle\langle \phi_{\mu}^{(s)}|$.
The vectors $|\phi_{\mu}^{(s)}\rangle$ denote the eigenvectors of $T_{\mu}$ associated with the eigenvalues $\lambda^{(s)}_{\mu}$.
Note that, generally speaking, the matrices $T_{\mu}$ cannot be simultaneously diagonalized.
We compute an expectation value of $T_{\mu}$ with a density matrix $\rho$ to obtain
\begin{align}
{\rm Tr}\left[ \rho T_{\mu} \right] 
&= \sum_{s=1}^{3} \lambda^{(s)}_{\mu} p^{(s)}_{\mu},  \\
p^{(s)}_{\mu} 
&= {\rm Tr}\left[ \rho |\phi_{\mu}^{(s)}\rangle\langle \phi_{\mu}^{(s)}| \right].
\end{align}
Following Ref.~\cite{Zhu19}, the coefficients $p^{(s)}_{\mu}$ are regarded as the probabilities for the discrete spins $d_{\mu} \in \{\lambda^{(1)}_{\mu},\lambda^{(2)}_{\mu},\lambda^{(3)}_{\mu}\}$.
In addition, we define the values of $\lambda^{(s)}_{\mu}$ as
\begin{gather}
\lambda^{(1)}_{\mu} = 1,\;\; \lambda^{(2)}_{\mu} = 0,\;\; \lambda^{(3)}_{\mu} = -1\;\;(\text{for}\;\; \mu \neq 8), \\
\lambda^{(1)}_{8} = \lambda^{(3)}_{8} = -\frac{1}{\sqrt{3}},\;\; \lambda^{(2)}_{8} = \frac{2}{\sqrt{3}},
\end{gather}
where $\sum_{s}\lambda^{(s)}_{\mu} = 0$ for all $\mu$.
Due to ${\rm Tr}\rho = 1$, each probability is normalized as $\sum_{s} p^{(s)}_{\mu} = 1$.
To combine this sampling scheme with the TWA, we assume that the equations of motion in Eq.~(\ref{eq: eom_dtwa}) are solved with initial conditions $x_{\mu}(t=0)=d_{\mu}$.

We determine the probabilities $p^{(s)}_{\mu}$ for the deep Mott-insulator state Eq.~(\ref{eq: density_matrix_sz_0}).
We see via direct computations that $d_{3}$, $d_{4}$, $d_{5}$, and $d_{8}$ do not fluctuate, because the density matrix of Eq.~(\ref{eq: density_matrix_sz_0}) leads to $p^{(2)}_{3} = p^{(2)}_{4} = p^{(2)}_{5} = p^{(2)}_{8} = 1$, and $p^{(s)}_{3} = p^{(s)}_{4} = p^{(s)}_{5} = p^{(s)}_{8} = 0$ for $s=1,3$.
However, the remaining ones, $d_1$, $d_2$, $d_6$, and $d_7$, can fluctuate: the nonzero probabilities for these are given by 
\begin{gather}
p^{(1)}_{1}=p^{(1)}_{2}=p^{(1)}_{6}=p^{(1)}_{7}=\frac{1}{2}, \nonumber \\
p^{(3)}_{1}=p^{(3)}_{2}=p^{(3)}_{6}=p^{(3)}_{7}= \frac{1}{2}.
\end{gather}
Therefore, in the TWA simulations, $d_{1}$, $d_{2}$, $d_{6}$, and $d_{7}$ randomly choose either $1$ or $-1$ with an equal probability, while $d_3=d_4=d_5=0$ and $d_8=2/\sqrt{3}$ for all samples.
Note that the fluctuations of each variable are statistically independent of those of the other ones.
More detailed discussions of the tomography technique are found in Ref.~\cite{Zhu19}.  
In Appendix~\ref{app: gdtwa}, we add a supplemental discussion on the relationship between this tomography method and our DTWA scheme, associated with the reproducibility of a second-order moment for a pure state.

\subsection{Fully connected spin-1 model} \label{sec:fully-connected}

To compare the SU(3)DTWA with the Gaussian SU(3)TWA, we study the fully connected spin-1 model (\ref{eq: fully-connected model}). 
To be specific, we calculate sudden-quench dynamics of several physical quantities by using the SU(3)TWA with the Gaussian-Wigner function and the SU(3)DTWA, respectively, and compare these semiclassical results with the exact ones.

\begin{figure*}
\includegraphics[width=180mm]{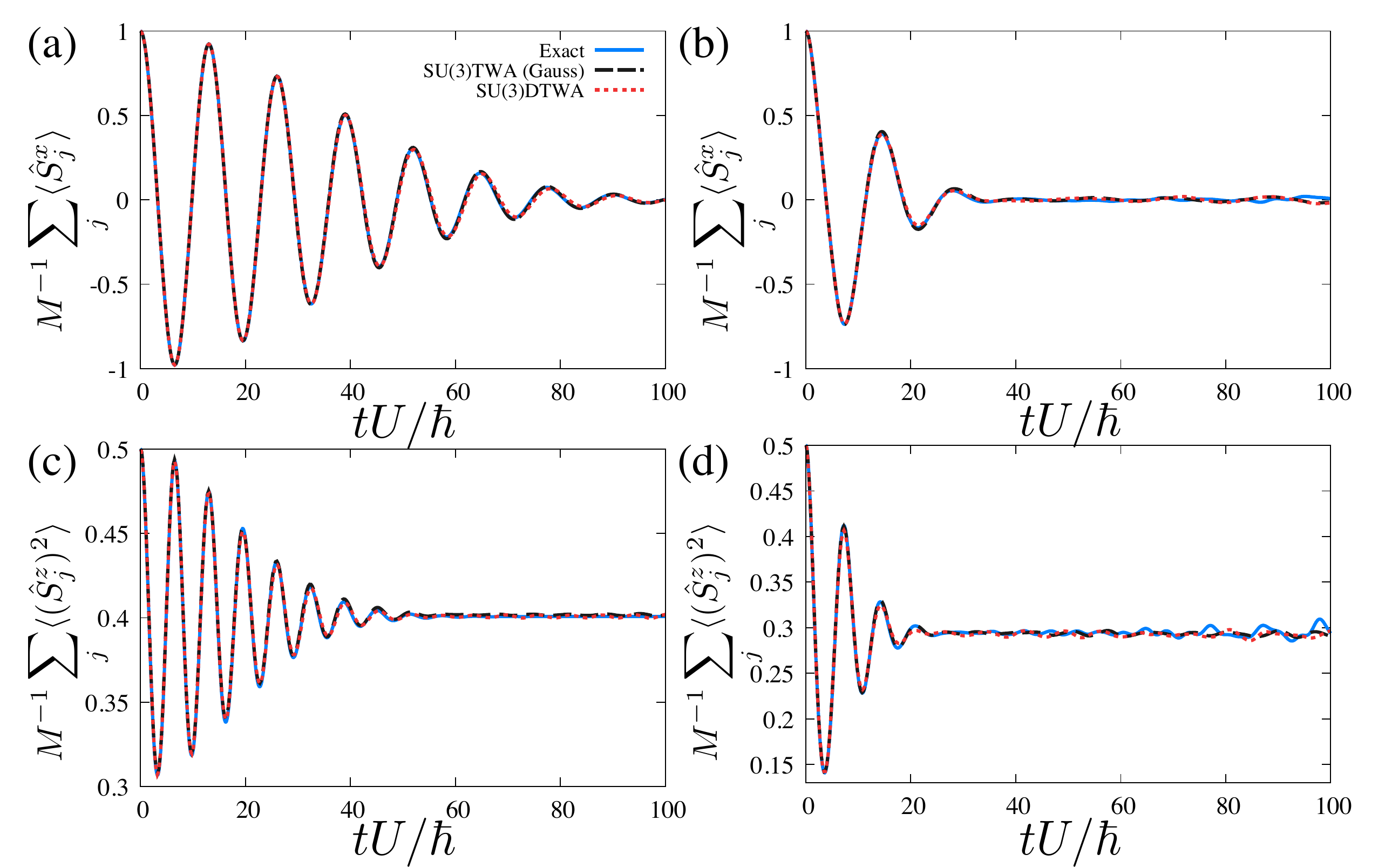}
\vspace{-3mm}
\caption{
Numerical simulation of the quench dynamics of (a), (b) $M^{-1}\sum_{j}\langle {\hat S}^{x}_{j} \rangle (t)$ and (c), (d) $M^{-1}\sum_{j}\langle ({\hat S}^{z}_{j})^2 \rangle(t)$ for the fully connected spin-1 model of Eq.~(\ref{eq: fully-connected model}) with $M=50$ sites. 
To initialize the system, we prepared the $x$-polarized state $|\Psi_0\rangle = \bigotimes_{j}|S^{x}_{j}=1\rangle$ at $t=0$. 
The black dashed and red dotted lines represent the results of the Gaussian SU(3)TWA and the SU(3)DTWA, respectively. 
The blue solid line means the exact quantum dynamics with the same initial condition. 
The left and right panels correspond to $U=250J \approx 5.1zJ$ and $125J \approx 2.6zJ$, respectively.
We note that the black dashed line in (c) reproduces well a panel of Fig. 2 in the previous work~\cite{Davidson15}.
}
\label{figure02}
\end{figure*}

In Fig.~\ref{figure02}, we numerically simulate the time evolution of the fully connected spin-1 model of Eq.~(\ref{eq: fully-connected model}) after sudden quenches from the $x$-polarized direct-product state
\begin{align}
|\Psi_0\rangle = \bigotimes_{j=1}^{M}|S^{x}_{j}=1\rangle. \label{eq:initial_state_x}
\end{align}
This state is a ground state of the model of Eq.~(\ref{eq: fully-connected model}) in the limit of $U/J\rightarrow 0$.
The number of lattice points is set $M=50$. 
For the simulation of the Gaussian SU(3)TWA, we numerically integrated the classical equation of motion for initial conditions, which are distributed by the Gauss probability distribution $P(\{X^{(j)}_{\mu}\})=\prod_{j=1}^{M}P_{j}(X^{(j)}_{1},\cdots,X^{(j)}_{8})$. 
The local distribution $P_{j}$ corresponds to the parameters in Eqs.~(\ref{eq:gauss_x_R}-\ref{eq:gauss_x_sigma}).
In the upper panels in Fig.~\ref{figure02}, we display the time evolution of $M^{-1}\sum_{j}\langle {\hat S}^{x}_{j}(t) \rangle$.
The semiclassical results of the Gaussian SU(3)TWA (black dashed line) agree quantitatively with the exact quantum dynamics (blue solid line) over a long timescale both for $U=250J$ ($U\approx 5.1 zJ$) [Fig.~\ref{figure02}(a)] and $U=125J$ ($U\approx 2.6 zJ$) [Fig.~\ref{figure02}(b)]. 
In the lower panels in Fig.~\ref{figure02}, we also show the time evolution of $M^{-1}\sum_{j}\langle ({\hat S}^{z}_{j})^2(t) \rangle$ starting from the same initial state.
It should be noticed that the slight recurrence of the oscillation observed in Fig.~\ref{figure02}(d) at late times after $t \approx 60\hbar/U$ are not captured within the semiclassical approximation as expected in typical TWA simulations \cite{Polkovnikov10,Davidson15}.

In Fig.~\ref{figure02}, we also simulate the same dynamics by using the SU(3)DTWA approach.
For all the panels, the SU(3)DTWA results (red dotted lines) reasonably reproduce the same dynamics as those of the Gaussian SU(3)TWA.
As explained in Appendix \ref{app:ppo}, for the DTWA results in Fig.~\ref{figure02}, we have prepared a statistical mixture of random initial conditions characterized by multiple sets of phase-point operators.
A similar technique has been used in Ref.~\cite{Pucci16}.
We emphasize that if we only use the Wootters representation for samplings, it will fail to correctly produce the dynamics [see also Fig.~\ref{figure_appendix_02}(a)]. 

\begin{figure}
\includegraphics[width=90mm]{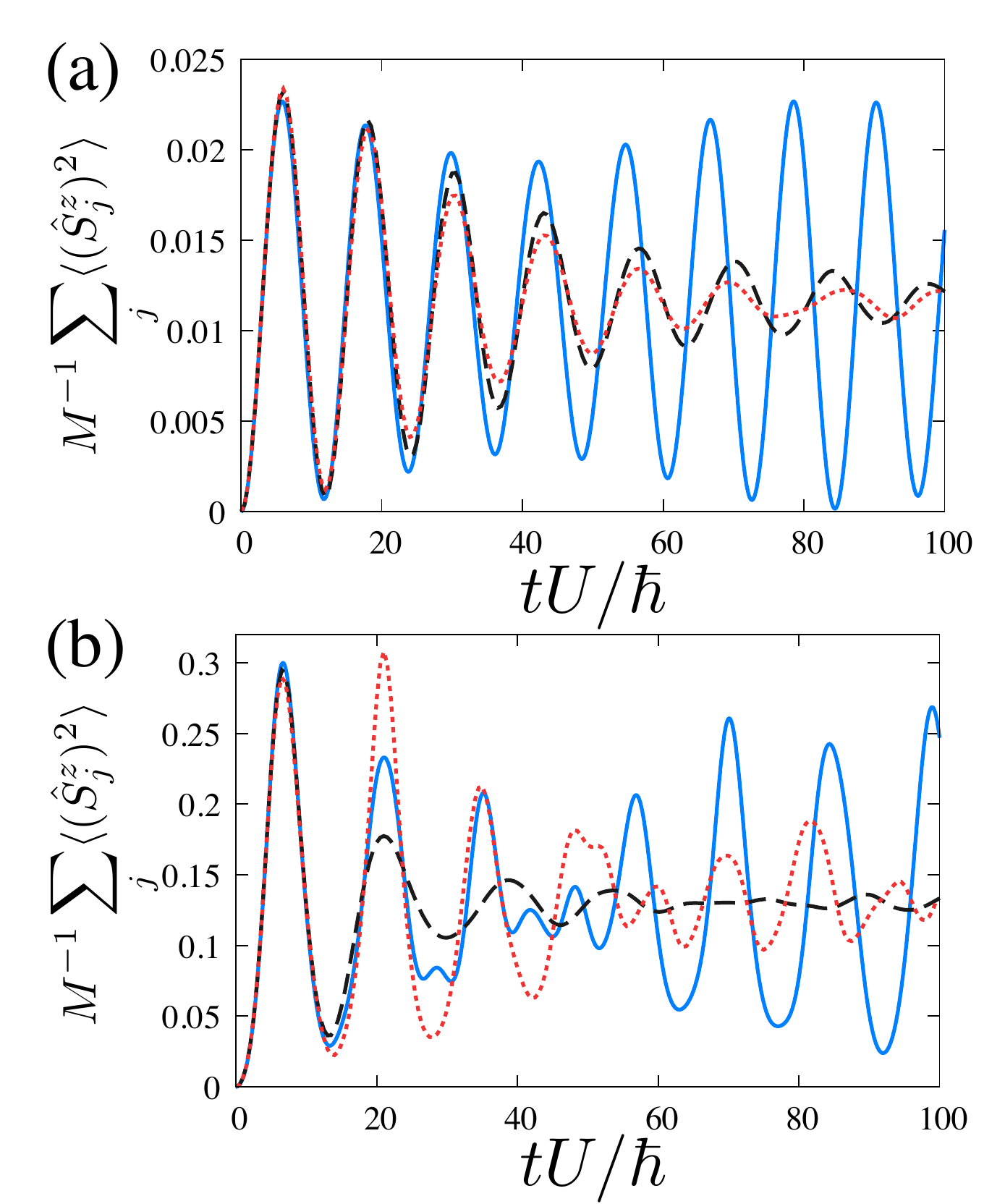}
\vspace{-5mm}
\caption{
Numerical simulation of the quench dynamics of $M^{-1}\sum_{j}\langle ({\hat S}^{z}_{j})^2 \rangle(t)$ for the fully connected spin-1 model of Eq.~(\ref{eq: fully-connected model}) at $M=50$, starting from the Mott-insulator state $|\Psi_{\rm Mott} \rangle = \bigotimes_{j=1}^{M}|S_{j}^{z}=0\rangle$. 
The upper (a) and lower (b) panels are obtained at $U=250J$ and $125J$, respectively. 
The black dashed, red dotted, and blue solid lines correspond to the Gaussian SU(3)TWA, SU(3)DTWA, and exact quantum dynamics, respectively.
For details of the SU(3)DTWA simulation of these panels, see also Appendix \ref{app:ppo}.
}
\label{figure03}
\end{figure}

In Fig.~\ref{figure03}, we compute the expectation value $M^{-1}\sum_{j}\langle ({\hat S}^{z}_{j})^2 \rangle(t)$ for another initial state 
\begin{align}
|\Psi_{\rm Mott} \rangle = \bigotimes_{j=1}^{M}|S_{j}^{z}=0\rangle.  \label{eq:zero-magnetization}
\end{align}
In the projected Hilbert space for the Bose-Hubbard model, this expresses the deep Mott-insulator state.
For a relatively large value of the onsite interaction, say $U=250J$ [Fig.~\ref{figure03}(a)], both Gaussian (black dashed) and discrete (red dotted) SU(3)TWA results reproduce the first and second peaks of the exact expectation value (blue solid) within $t<30\hbar/U$.
As $U/J$ decreases, the timescale, during which the exact quantum dynamics are reasonably captured by the semiclassical expressions, is shortened.
This tendency can be attributed to the non-linearity of the system that gives rise to a significant error in the exact time evolution of the many-body Wigner function.
In Fig.~\ref{figure03}(b) corresponding to $U=125J$, both semiclassical approaches only recover the first peak within $t<10\hbar/U$, however, they fail to describe the second peak, especially, its amplitude.

After leaving from the early-time stage, the SU(3)TWA clearly deviates from the exact dynamics.
In particular, it is clearly seen in Fig.~\ref{figure03} that the Gaussian SU(3)TWA tends to saturate into a steady value but not to make a recurrence of the oscillation, both for $U=250J$ and $125J$.
Interestingly, especially in Fig.~\ref{figure03}(b), while the SU(3)DTWA also fails to describe the exact dynamics for $t > 10\hbar/U$, but it exhibits an oscillatory behavior rather than saturation. 
However, it should be emphasized that the discrete Monte Carlo sampling does not affect the quantitative timescale itself, during which the quantum dynamics are almost accurately captured within the semiclassical expressions.
This seems to be reasonable because the classical equations of motion for the continuous and discrete cases are the same.

To close this section, we have demonstrated that the SU(3)DTWA is nearly as accurate as the Gaussian approximation with respect to simulating the quench dynamics.
In the next section, we apply these techniques to analyses of the experimental results for 2D Bose-Hubbard systems \cite{Takasu20}.

\section{Application to the 2D Bose-Hubbard system} \label{sec:5}

In this section, we apply the SU(3)TWA approaches for studying far-from-equilibrium dynamics of the Bose-Hubbard model on a square lattice at unit filling. 
We specifically analyze dynamics of equal-time single-particle correlation functions after a quench from a Mott-insulating state to a parameter region near the quantum critical point \cite{Takasu20}. 
Theoretical studies on dynamics of equal-time correlation functions have been reported in Refs.~\cite{Lauchli08,Mathey10,Barmettler12,Natu13,Carleo14,Bonnes14,Krutitsky14,Richerme14,Schachenmayer15Dynamics,Fitzpatrick18,Nagao19}.

\subsection{Experimental setup}

First we briefly summarize the details of the experimental setup in Ref.~\cite{Takasu20}. 
Takasu and his coworkers have measured sudden-quench dynamics of the single-particle correlation functions inside the 2D Mott-insulator phase in the following steps: 
\begin{enumerate}
\item{
They prepared a unit-filling Mott insulator of an ultracold $^{174}{\rm Yb}$ gas in an optical square lattice with $s=V_0/E_{\rm R}=15$. 
The energy scales $V_0$ and $E_{\rm R}$ denote the optical lattice depth and the recoil energy of this system, respectively. 
The prepared system is well described by the direct product Fock state for bosons
\begin{align}
|\Psi_{\rm ini} \rangle \approx \bigotimes_{j=1}^{M}| n_{j} = 1 \rangle,
\end{align}
where ${\hat n}_{j}| n_{j} \rangle = n_{j} | n_{j} \rangle$.
}
\item{
The lattice depth was abruptly decreased from $s=15$ to $s=9$. 
The time to ramp down the lattice depth is approximately 0.1 ms. The lattice depth after the quench implies $U/J=19.6$.
}
\item{
After the quench, the resulting dynamics was observed by measuring the time-of-flight interference pattern that can be converted to the equal-time single-particle correlation functions,
\begin{align}
K_{{\boldsymbol \varDelta}}(t)=\frac{1}{M{\bar n}}\sum'_{{\bm r}_j,{\bm r}_{j'}} \langle {\hat a}^{\dagger}_{j}(t){\hat a}_{j'}(t) \rangle,
\end{align}
where ${\bm r}_{j}=(x_j,y_j)$ indicates each site on the square lattice with units of the lattice constant $d_{\rm lat}=266\;{\rm nm}$.
The real-space summation is performed under the conditions $|x_{j}-x_{j'}| = \varDelta_x$ and $|y_{j}-y_{j'}| = \varDelta_y$, and we write ${\boldsymbol \varDelta}=(\varDelta_x,\varDelta_y)$.
Recall that $M$ is the total number of lattice points. 
}
\end{enumerate}
In this work, as a simplified setup, we neglect harmonic trap potentials in numerical simulations. 
We simply assume that all the atoms participate in a uniform Mott-insulator state before the quench.
Effects due to spatial inhomogeneity of the gases will be discussed in Sec.~\ref{sec:discussions}.

To close this subsection, here we note that the qualitative behaviors of the dynamics of the spatial correlation functions measured after quantum quenches can change depending on the initial states that we take.
For instance, for the coherent state as the initial states, which describes a coherent condensation of bosons at the non-interacting limit, sudden changes of the interaction, from zero to weak interactions, result in observing fine oscillations in time of the density-density equal-time correlation function, reflecting the coherent motion of the Bogoliubov quasiparticles~\cite{Nagao19}. 
By contrast, if we choose the Mott-insulator states as the initial conditions, and propagate the states with the Hamiltonian with the same interactions (i.e., quenches from infinite to weak interactions), we observe propagation of a peak signal without fine oscillations in the same correlation function~\cite{Nagao19}.
Its propagation velocity is well explained by the single-particle excitation spectrum of the Hartree-Fock approximation. 
Reliable TWA results on this kind of initial-state dependence of the quench dynamics can be found in our previous study for the 2D Bose-Hubbard model with a large filling factor~\cite{Nagao19}.

\subsection{Small size case} \label{sec:small_size}

Before proceeding to our main results corresponding to the experimental setup, let us consider the quench dynamics for a small-size 2D Bose-Hubbard system, say 9 sites, in order to compare outputs of the SU(3)TWA approaches with those of the exact numerical calculation. 
For simplicity, we focus on the sudden-quench limit, in which the ramp-down time is neglected.

\begin{figure}
\begin{center}
\includegraphics[width=90mm]{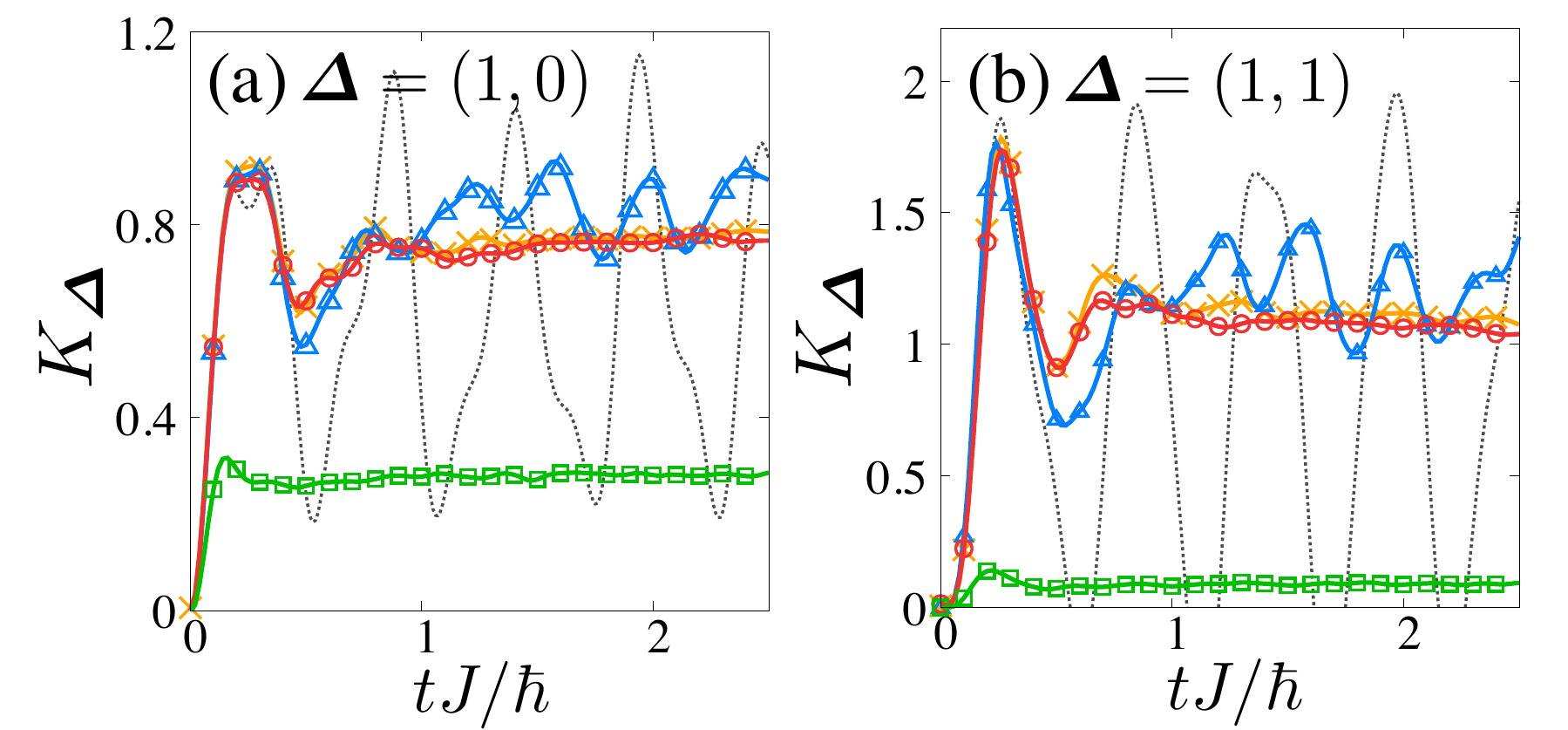}
\vspace{-5mm}
\caption{Time evolution of the correlation function $K_{{\boldsymbol \varDelta}}(t)$ of the Bose-Hubbard model on a square lattice for $M=3^2=9$ with periodic boundary conditions. 
We set the maximum occupation of particles per site $n_{\rm max}$ to be 2.
The interaction after the quench is $U/J=19.6$, corresponding to the experiment. 
The solid lines with points are the semiclassical results: Gaussian SU(3)TWA (red circle); SU(3)DTWA (blue triangle); GPTWA (green square); and the tomography sampling method indicated in Sec.~\ref{sec:dtwa} (orange cross). 
The gray dotted line represents the exact quantum dynamics. 
The left and right panels correspond to ${\boldsymbol \varDelta}=(1,0)$ and $(1,1)$, respectively.}
\label{figure04}
\end{center}
\end{figure}

Figure~\ref{figure04} shows the numerical results for the sudden-quench dynamics of $K_{\boldsymbol \varDelta}(t)$ for a small-size Bose-Hubbard system. 
The simulation setup has $M=3^2=9$ sites and we adopt periodic boundary conditions \cite{footnote}.
In Fig.~\ref{figure04}, the full-quantum dynamics of the Bose-Hubbard system is evaluated by integrating the time-dependent Schr\"odinger equation of the Hamiltonian (\ref{eq: bhm}) (gray dotted line).
The maximum occupation of the local site is $n_{\rm max}=2$, hence, the three lowest states, i.e., $|0\rangle$, $|1\rangle$, $|2\rangle$, are allowed in this simulation.
We observe that the correlation functions at ${\boldsymbol \varDelta}=(1,0)$ and $(1,1)$ form a first-peak region in the time range of $0 < tJ/\hbar < 0.5$.
At later times, $tJ/\hbar >0.5$, the time evolution of correlations exhibits an almost undamped oscillation reflecting its small size.

In Fig.~\ref{figure04}, we also simulate the same dynamics by using the SU(3)TWA for the effective-model Hamiltonian (\ref{eq:effective-model-low}) according to the Gaussian and discrete-Wigner approaches of Monte Carlo samplings. 
The unit-filling Mott-insulator state is given by Eq.~(\ref{eq:zero-magnetization}), i.e., $|\Psi_{\rm ini} \rangle \approx |\Psi_{\rm Mott} \rangle$.
We observe that both the Gaussian SU(3)TWA (red circle) and SU(3)DTWA (blue triangle) quantitatively capture the first-peak region in the range of $0 < tJ/\hbar < 0.5$, especially its initial growth, its time point of the center of the region, and its correlation intensity. 
However, the later-time dynamics for $tJ/\hbar > 0.5$ can not be well captured within the SU(3) semiclassical representation.
Indeed, the semiclassical results exhibit almost saturated behaviors rather than the temporal oscillation with a large amplitude.
It should be emphasized that the difference between two semiclassical results in the later-time dynamics comes from our choice of the initial distribution for the phase-space variables.
Interestingly, it is clearly seen that, around $t = 0.5\hbar/J$ in Fig.~\ref{figure04}, the SU(3)DTWA gives a slightly better result, i.e., shows deeper dips of correlations.
For this comparison, the SU(3)DTWA can be seen as a better description than the Gaussian SU(3)TWA.

Figure~\ref{figure04} also displays the simulation result on the basis of the tomography technique as presented in Sec.~\ref{sec:dtwa}. 
We numerically find that it is closer to the Gaussian result, rather than the DTWA one.
This coincidence to the Gaussian simulation indicates that the tomography technique also provides a reasonable sampling scheme for the initial condition. 
Since there is no considerable deviation from the Gaussian result, in the following discussions, we do not use the tomography technique.

It is interesting and helpful to calculate the quench dynamics by using the GPTWA for the strongly interacting Bose-Hubbard system as a reference. 
In order to carry out an efficient simulation, we have used an approximate Gaussian distribution representing the Fock states~\cite{Nagao19}. 
The details of the GPTWA will be briefly reviewed in Appendix \ref{app:twa}. 
In Fig.~\ref{figure04}, the GPTWA simulation (green square) fails to describe the correlation intensity in the first-peak region while it reproduces well a very early growth of the correlation function at ${\boldsymbol \varDelta}=(1,0)$ within $tJ/\hbar < 0.2$.
Therefore, for the purpose of simulating the strongly interacting dynamics, the SU(3)TWA certainly provides a better description than the GPTWA.

\subsection{Comparison to the experimental results}

\begin{figure*}
\includegraphics[width=170mm]{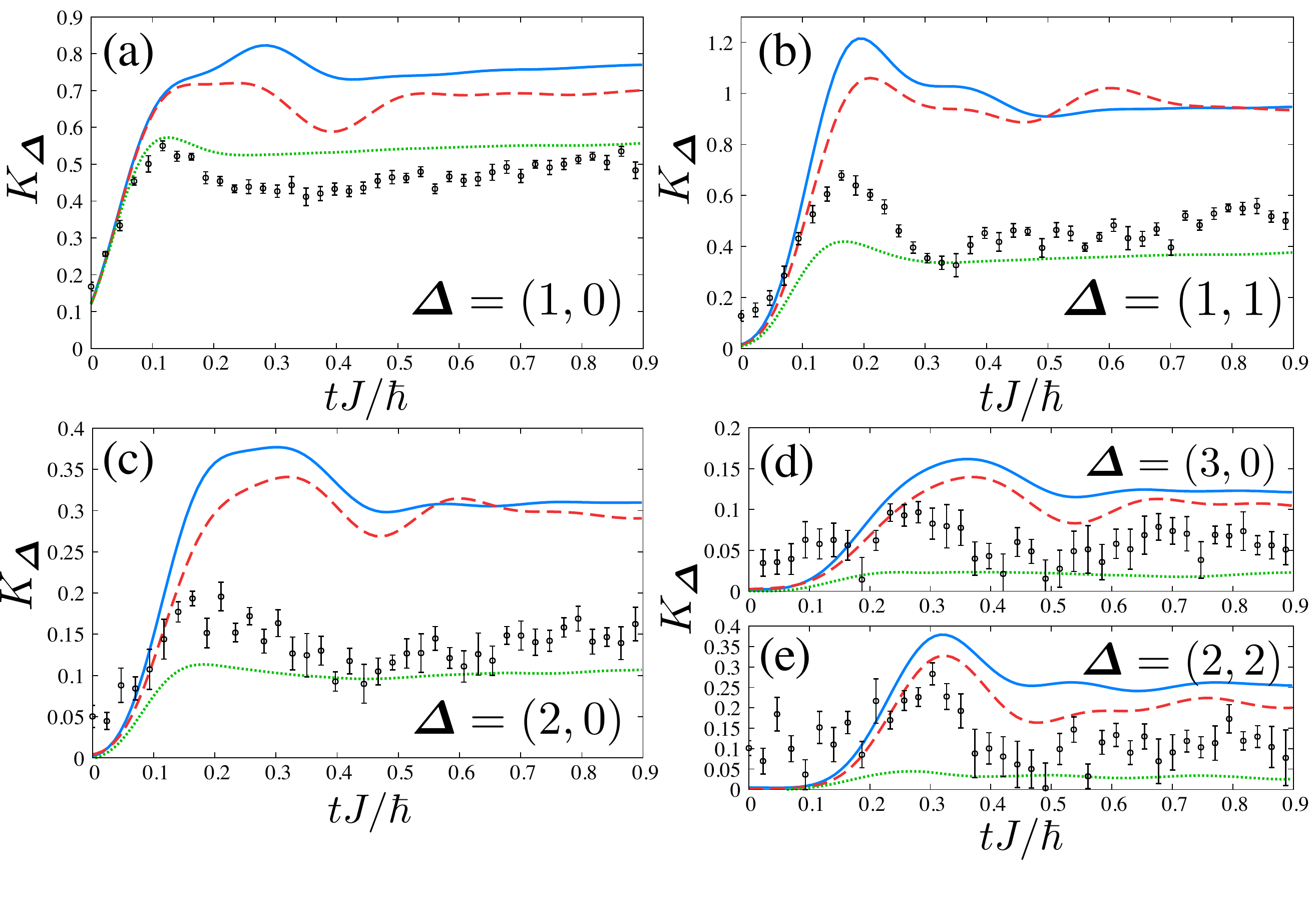}
\vspace{-5mm}
\caption{
Time evolution of $K_{\boldsymbol \varDelta}(t)$ for the 2D Bose-Hubbard model with $n_{\rm max} = 2$ and periodic boundary conditions ($M=20^2=400$).
The initial state is the Mott-insulator state.
The finite ramp-down time has been taken into account in this simulation.
During the time evolution for $t>0$, the system has the interaction strength $U/J=19.6$.
The black points with error bars express the experimental results in Ref.~\cite{Takasu20}.
The green dotted lines (GPTWA) are a reproduction from the same reference.
The red dashed and blue solid lines are the Gaussian SU(3)TWA and the SU(3)DTWA, respectively.
For the SU(3)TWA (GPTWA), we have taken $4000$ ($6000$) trajectories. 
}
\label{figure05}
\end{figure*}

We calculate the quench dynamics for a larger-size system corresponding to the experimental setup. 
Figure~\ref{figure05} shows the correlation function $K_{\boldsymbol \varDelta}(t)$ for $M=20^2 = 400$ with periodic boundary conditions.
First, we prepare the system in the unit-filling Mott-insulator state ($t<0$), and then abruptly decrease the lattice depth until $t=0$.
For $t > 0$, the system evolves in time at $U=19.6J$.
While the dynamics of the effective pseudospin-1 model (\ref{eq:effective-model-low}) is computed in the SU(3)TWA simulations, that of the Bose-Hubbard model (\ref{eq: bhm}) with no truncation of the local Hilbert space is computed in the GPTWA.
We note that the GPTWA result in Fig.~\ref{figure05} is a reproduction from Ref.~\cite{Takasu20}.

In Fig.~\ref{figure05}(a), we observe that all the semiclassical results explain well the growth of the nearest-neighbor correlation at ${\boldsymbol \varDelta}=(1,0)$ in the early-time stage within $t < 0.1\hbar/J$. 
In addition, these reasonably describe the correlation offset at $t=0$.
The experimental data show a peak in the time domain of $0 < t J/\hbar < 0.2$. 
At longer times, the measured correlation gradually saturates to a steady value.  
In the comparison performed in Fig.~\ref{figure05}(a), the experimental result is seemingly closer to the GPTWA rather than the SU(3)TWA.
In particular, the peak position and the correlation intensity in the time window indicated by Fig.~\ref{figure05}(a) are relatively close to the ones simulated by the GPTWA.
This is in contrast to the small-size case in Sec.~\ref{sec:small_size}, where the SU(3)TWA is closer to the exact dynamics and can provide a reasonable first-peak region at short times.
Notice that the correlation intensity of the experiment is typically lesser than both SU(3)TWA and GPTWA results.

Next, we focus on longer distances, say ${\boldsymbol \varDelta}=(1,1)$ [Fig.~\ref{figure05}(b)] and ${\boldsymbol \varDelta}=(2,0)$ [Fig.~\ref{figure05}(c)].
The experimental data are seen to achieve a peak during $0.1 < tJ/\hbar < 0.3$ for both cases. 
The time points of the center of the first-peak regions are reasonably captured by both SU(3)TWA and GPTWA, at least within the accuracy of experimental errors.
In Figs.~\ref{figure05}(b) and (c), the GPTWA typically produces smaller values of correlations compared with the experimental result. 
Likewise in the case of ${\boldsymbol \varDelta}=(1,0)$, the SU(3)TWA tends to yield larger values than those of the experiment.
Again, the experiment is seemingly closer to the GPTWA rather than the SU(3)TWA.

For further longer distances, say ${\boldsymbol \varDelta}=(3,0)$ [Fig.~\ref{figure05}(d)] and ${\boldsymbol \varDelta}=(2,2)$ [Fig.~\ref{figure05}(e)], it is hard to locate the center of the first-peak region in the experimental data because of significant noises.
Within the error bars, we expect that there exists a peak region somewhere in the range of $t < 0.5 \hbar/J$.
For these long distances, correlation intensities in the GPTWA are seen to be suppressed because it cannot capture strong quantum fluctuations in the parameter regime.
In particular, no clear peak region is observed in the simulation even at short times.
Therefore, its agreement to the experiment is worse.
By contrast, the SU(3)TWA, which is expected to describe local quantum fluctuations in the regime more accurately, can produce a reasonably strong correlation, which is comparable to the experiment, and clear peak regions in the range of $t < 0.5 \hbar/J$.
Hence, in this case, these SU(3) simulations are closer to the experiment.

Finally, let us mention that, in the experiment, not only the nearest-neighbor correlation but also the longer-distance ones exhibit a finite and non-negligible offset at $t=0$.
However, according to the SU(3)TWA and the GPTWA, such an offset for longer distances should be almost zero.
We will discuss this point in details in the next section. 

\subsection{Discussions} \label{sec:discussions}

In the direct comparisons for the large system, we observed that the experimental results of the spatial correlation function are closer to the GPTWA especially at short distances while the SU(3)TWA looks better at long distances.
As learned from the numerical simulations for the small size, the SU(3)TWA should work better more than the GPTWA in the strongly interacting parameter regime.
Moreover, we also recognized that the nonzero offsets of the correlations at distances except for nearest neighbors are not consistent to all the semiclassical results.
We argue that the above unexpected observations could be attributed to some contributions present in the actual experiment, which are not precisely taken into account in our SU(3)TWA simulations.

First, we discuss occupations of bosons allowed in the SU(3)TWA for the Bose-Hubbard systems. 
The formalism of SU(3)TWA for bosons is constructed under assumptions that the local Hilbert space is truncated up to three states.
If the dimension of the reduced state space is extended from three to more, it will improve the simulated result, more or less, quantitatively.
To perform this extension, one needs to increase the local phase space furthermore.
For instance, if five states are relevant locally, SU(5) matrices should be chosen as a phase-space variable.
However, we may expect that higher occupations give no significant effect, at least in our current case, in which the strength of the interaction is large enough to suppress them.
In order to justify this expectation, in Appendix~\ref{app: supplemental}, we will clarify the degree to which occupations greater than 2 affect the quench dynamics of the correlation function in the parameter regime of the experiment by utilizing an exact numerical calculation for a small size.

Second, we make a comment on effects of an inhomogeneous trap potential.
In the experimental setup in Ref.~\cite{Takasu20}, the prepared initial state actually contains a strongly correlated superfluid component with incommensurate fillings due to a harmonic trap while the region of the Mott insulator with unit filling is much larger.
Such a contribution is not dominant over the whole gas, but not completely negligible.
In the Supplemental Material of Ref.~\cite{Takasu20}, an MPS calculation has been performed for a 1D trapped Bose gas in the presence of narrow superfluid regions in the system.
A numerical result shows that a finite offset appears at the end point of quenches at several distances in addition to the nearest neighbor. 
This strongly indicates that the presence of superfluid contributions, more or less, affects the time evolution of the correlation function. 
To our current techniques for the SU(3)TWA, it is difficult to initialize a system into such inhomogeneous states as prepared in the MPS simulation.
In future works we will develop an efficient technique to treat this kind of initialization problem.

\section{Conclusions and outlooks} \label{sec:6}

In conclusion, we have analyzed far-from-equilibrium dynamics of strongly interacting Bose gases in an optical lattice by using the SU(3)TWA on the basis of different Monte Carlo sampling schemes.
In the middle of this paper (Sec.~\ref{sec:4}), the SU(3)DTWA approach has been developed as a sampling scheme, and applied to the fully connected spin-1 model with a large size in order to examine this approach.
We demonstrated that the SU(3)DTWA is nearly as accurate as the Gaussian SU(3)TWA in simulating time evolution after sudden quantum quenches.
  
In the main part of this paper (Sec.~\ref{sec:5}), we have applied the SU(3)TWA to sudden quench dynamics of a strongly interacting Bose gas in the 2D optical lattice.
The semiclassical methods on the basis of the GPTWA, the SU(3)DTWA, and the Gaussian SU(3)TWA have been compared with exact numerical calculations for the 2D Bose-Hubbard model with a small size. 
We recognized that the SU(3)DTWA and the Gaussian SU(3)TWA can provide better descriptions than the GPTWA in a strongly interacting regime.
The numerical results on the basis of those semiclassical methods have also been compared with the recent experiment at Kyoto University.
We found that at short distances, the experiment is closer to the GPTWA while, at relatively-long distances, it is reasonably close to the SU(3)DTWA and the Gaussian SU(3)TWA.
We argued that this observation can be attributed to parts of the actual experimental realization including an inhomogeneous trap potential, which are not precisely taken into account in our numerical simulations.

Beyond the scope of this work, it would be interesting to develop a cluster TWA approach \cite{Wurtz18} for the strongly interacting Bose-Hubbard systems.
For applications of this strategy in higher dimensions than 1D, a reasonable reduction scheme of dimensions of cluster phase-space variables may be required to make simulations realistic and efficient.

\begin{acknowledgments}

We thank A. Polkovnikov, M. Kunimi, and S. Goto for useful discussions. We also thank S. Davidson for providing insightful comments on the numerical simulation of the fully connected spin-1 model. This work was supported by KAKENHI from Japan Society for the Promotion of Science (Grants No.~JP18K03492, No.~JP18H05228, No.~JP25220711, No.~JP17H06138, No.~JP18H05405, and No.~JP16H00801), the Impulsing Paradigm Change through Disruptive Technologies (ImPACT) program, CREST from Japan Science and Technology Agency Grant No.~JPMJCR1673, MEXT Quantum Leap Flagship Program (Q-LEAP) Grant No. JPMXS0118069021, and the Matsuo Foundation. 

\end{acknowledgments}

\appendix

\section{Exact numerical dynamics of the fully connected spin-1 model} \label{app:diag}

The large-size numerical simulation of the exact dynamics of the fully connected spin-1 model is carried out as follows. 
Let us begin with introducing collective SU(3) operators defined by 
\begin{align}
{\hat \Pi}_{\alpha} = \sum_{j=1}^{M}{\hat X}^{(j)}_{\alpha},\;\;\;{\rm for}\; \alpha = 1,2,\cdots,8.
\end{align}
From the properties of the local SU(3) generators ${\hat X}^{(j)}_{\alpha}$, ${\hat \Pi}_{\alpha}$ should satisfy the commutation relation of the SU(3) group 
\begin{align}
[{\hat \Pi}_{\alpha}, {\hat \Pi}_{\beta}] = if_{\alpha \beta \gamma}{\hat \Pi}_{\gamma}.
\end{align}
Then, the fully connected model can be expressed as
\begin{align}
{\hat H}_{\rm fc} = -\frac{J}{2}\left({\hat \Pi}_{1}^2 + {\hat \Pi}_{2}^2 \right) + \frac{1}{2\sqrt{3}}(J-U){\hat \Pi}_{8}.
\end{align}
The eigenstate of ${\hat \Pi}_{3}$ characterizes the global spin polarization over the whole system. 
For example, the maximally polarized state of ${\hat \Pi}_{3}$ is equal to the state, in which all of ${\hat X}^{(j)}_3$ at each site are entirely polarized along $z$ axis, i.e., $|1111\cdots \rangle$. 

As the next step, we rewrite the collective operators in the form of the bilinear bosonic operator through the Jordan-Schwinger mapping \cite{Auerbach12}. 
Then, by using the SU(3) Schwinger boson $({\hat b}_{1},{\hat b}_0,{\hat b}_{\bar 1})$, the operator ${\hat \Pi}_{\alpha}$ takes the form
\begin{align}
{\hat \Pi}_{\alpha}=\sum_{m,n}{\hat b}^{\dagger}_{m}T^{mn}_{\alpha}{\hat b}_{n},\;\;\; \sum_{n} {\hat b}^{\dagger}_{n}{\hat b}_{n} = M {\hat 1}.
\end{align}
Notice that the bosons are constrained by $M$, but not unity. 
This expression may be regarded as an SU(3) analog of the Schwinger-boson representation of the SU(2) generators characterized by a spin strength $S$ \cite{Auerbach12}.

Subject to a fixed $M$, an arbitrary state of this system is spanned by a Fock vector labeled by two non-zero integers $\nu_1 \geq 0$ and $\nu_2 \geq 0$,
\begin{align}
|\nu_1,\nu_2\rangle = \frac{({\hat b}^{\dagger}_{1})^{\nu_1}({\hat b}^{\dagger}_{0})^{M-\nu_1-\nu_2}({\hat b}^{\dagger}_{\bar 1})^{\nu_2}}{\sqrt{\nu_1!}\sqrt{\nu_2!}\sqrt{(M-\nu_1-\nu_2)!}}|{\rm vac}\rangle,
\end{align}
where $0 \leq \nu_1 + \nu_2 \leq M$. 
This basis state is a simultaneous eigenstate for ${\hat \Pi}_{3}$ and ${\hat \Pi}_{8}$, therefore,
\begin{align}
{\hat \Pi}_{3}|\nu_1,\nu_2\rangle &= (\nu_1 - \nu_2)|\nu_1,\nu_2\rangle,\nonumber \\
{\hat \Pi}_{8}|\nu_1,\nu_2\rangle &= \left[ \frac{2M}{\sqrt{3}} - \sqrt{3}(\nu_1 + \nu_2) \right] |\nu_1,\nu_2\rangle.
\end{align}
The rest of the operators, e.g., ${\hat \Pi}_{1}$, behave as a ladder operator connecting different Fock states
\begin{align}
{\hat \Pi}_{1}|\nu_1,\nu_2\rangle &= \frac{1}{\sqrt{2}}{\hat b}^{\dagger}_{1}{\hat b}_{0}|\nu_1,\nu_2 \rangle + \frac{1}{\sqrt{2}}{\hat b}^{\dagger}_{0}{\hat b}_{1}|\nu_1,\nu_2 \rangle \nonumber \\
&\;\;\;\;\;\; +\frac{1}{\sqrt{2}}{\hat b}^{\dagger}_{\bar 1}{\hat b}_{0}|\nu_1,\nu_2 \rangle + \frac{1}{\sqrt{2}}{\hat b}^{\dagger}_{0}{\hat b}_{\bar 1}|\nu_1,\nu_2 \rangle \nonumber \\
&=\frac{1}{\sqrt{2}}\sqrt{(M-\nu_1-\nu_2)(\nu_1+1)}|\nu_1+1,\nu_2\rangle \nonumber \\
&+ \frac{1}{\sqrt{2}}\sqrt{\nu_1(M-\nu_1-\nu_2+1)}|\nu_1-1,\nu_2\rangle  \nonumber \\
&+ \frac{1}{\sqrt{2}}\sqrt{(M-\nu_1-\nu_2)(\nu_2+1)}|\nu_1,\nu_2+1\rangle \nonumber \\
&+ \frac{1}{\sqrt{2}}\sqrt{\nu_2(M-\nu_1-\nu_2+1)}|\nu_1,\nu_2-1\rangle. \nonumber
\end{align}
One can evaluate the matrix element of the Hamiltonian between $|\nu_1,\nu_2\rangle$ and $|\nu_1',\nu_2' \rangle$, i.e., $\langle \nu_1',\nu_2' | {\hat H}_{\rm fc} | \nu_1,\nu_2 \rangle$. 
Its dimension algebraically increases with $M$, so that one can implement the exact numerical analysis on computers even at large $M$.

\begin{figure}
\vspace{-5mm}
\includegraphics[width=85mm]{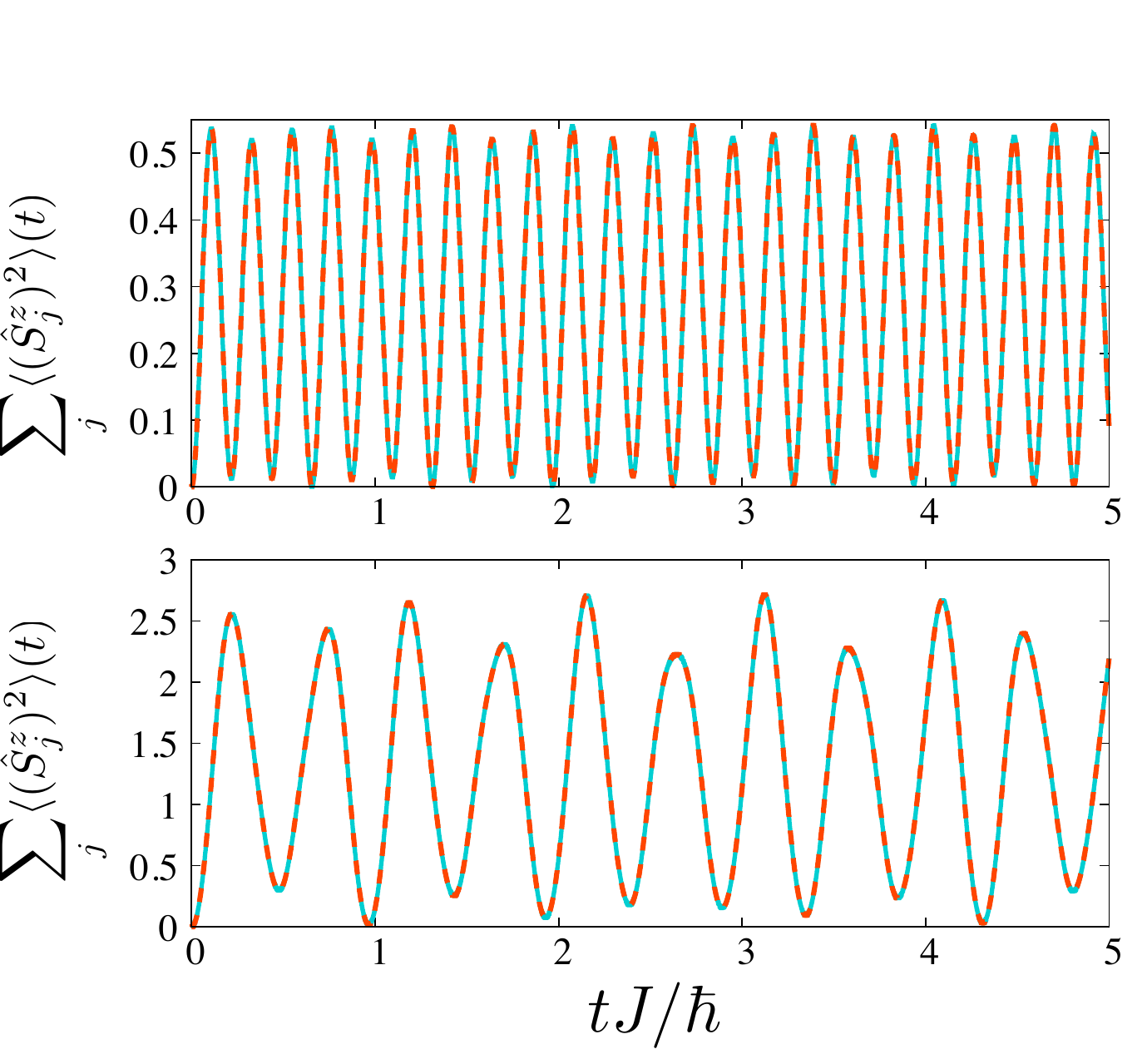}
\vspace{-2mm}
\caption{Exact quantum dynamics of $\sum_{j}\langle ({\hat S}^{z}_{j})^2 \rangle(t)$ for the fully connected spin-1 model for $M=8$. 
The dashed line is the result for the collective-spin representation of the Hamiltonian. 
On the other hand, the solid line is obtained in the original spin expression of the Hamiltonian, whose Hilbert-space dimension exponentially increases with $M$. 
The top and bottom panels correspond to $U=40J$ and $20J$, respectively.}
\label{figure_app_01}
\end{figure}

In Fig.~\ref{figure_app_01}, we compute the time evolution of the expectation value $\sum_{j}\langle ({\hat S}^{z}_{j})^2 \rangle(t)=2M/3-\langle {\hat \Pi}_8(t)\rangle/\sqrt{3}$ for $M=8$ using two different approaches: the dashed line is a numerical integration of the time-dependent Schr\"odinger equation for the Hamiltonian matrix expressed in terms of the collective spin ${\hat \Pi}_{\alpha}$, whereas the solid line is for the spin Hamiltonian in terms of ${\hat S}^{\alpha}_{j}$. 
The initial state for this simulation is the zero-magnetization direct-product state $|\Psi(t=0)\rangle = \bigotimes_{j}|S^{z}_{j}=0\rangle$. 
The perfect agreement of two results manifests that the collective-spin expression gives a more efficient way to have the same result than the straightforward approach.

\section{Gross-Pitaevskii truncated-Wigner approximation for the Bose-Hubbard Hamiltonian} \label{app:twa}

For the TWA in the coherent-state phase space, the classical time evolution of the Bose-Hubbard Hamiltonian is governed by the discrete GP equation associated with the Heisenberg-Weyl group:
\begin{align}
i\hbar \frac{\partial \alpha_{j}}{\partial t} = \frac{\partial H_{W}}{\partial \alpha^*_{j}},\;\; i\hbar \frac{\partial \alpha^*_{j}}{\partial t} = -\frac{\partial H_{W}}{\partial \alpha_{j}}. \label{app:eq:gpe}
\end{align}
The classical function $H_{W}(\bm{\alpha},\bm{\alpha}^*)=({\hat H}_{\rm BH})_{W}$ is the Weyl symbol of ${\hat H}_{\rm BH}$ given by
\begin{align}
H_{W} &= -J\sum_{\langle i,j \rangle}(\alpha^*_{i}\alpha_{j}+\alpha^*_{j}\alpha_{i})  \nonumber \\
&\;\;\;\;\;\;\;\; + \frac{U}{2}\sum_{j}\left[|\alpha_{j}|^4 - 2|\alpha_{j}|^2 + \frac{1}{2} \right]. \label{app:eq:hw}
\end{align}
If we write $\bm{\alpha}_{\rm cl}(t)$ as a solution of the GP equation with conditions $\bm{\alpha}_{\rm cl}(t=0)=\bm{\alpha}_0$, the expectation value of an operator ${\hat \Omega}$, i.e., $\langle {\hat \Omega}(t) \rangle = {\rm Tr}[{\hat \Omega}{\hat \rho}(t)] = {\rm Tr}[{\hat \Omega}(t){\hat \rho}(t=0)]$ is reduced to the following phase-space integration form (for details, see \cite{Nagao19,Polkovnikov10,Blakie08}):
\begin{align}
\langle {\hat \Omega}(t) \rangle \approx \int d\bm{\alpha}_0d\bm{\alpha}^*_0 \Omega_{W}[\bm{\alpha}_{\rm cl}(t),\bm{\alpha}^*_{\rm cl}(t)]W(\bm{\alpha}_0,\bm{\alpha}^*_0).
\end{align}
Here $d\bm{\alpha}d\bm{\alpha}^* = \pi^{-M}\prod_{j=1}^{M}d{\rm Re}[\alpha_{j}]d{\rm Im}[\alpha_{j}]$ is the measure of the phase-space integration. The weight function over the phase space is the Wigner function defined by means of the coherent state basis
\begin{align}
W(\bm{\alpha},\bm{\alpha}^*) 
&= ({\hat \rho}(t=0))_{W} \nonumber \\
&= \int \frac{d\bm{\eta}d\bm{\eta}^*}{2^{M}} \left\langle \bm{\alpha} - \frac{\bm{\eta}}{2} \right| {\hat \rho}(t=0) \left| \bm{\alpha} + \frac{\bm{\eta}}{2} \right\rangle \nonumber \\
&\;\;\;\;\;\; \times e^{\frac{1}{2}(\bm{\eta}^*\cdot \bm{\alpha} - \bm{\eta}\cdot \bm{\alpha}^*)  }.
\end{align}

We note that the GPTWA typically provides quantitative descriptions of real-time dynamics of the Bose-Hubbard systems when they have a sufficiently small interaction or sufficiently large filling factor. 
In recent years, this type of semiclassical method has been applied to multiple dynamical problems of lattice bosons, e.g., see Refs.~\cite{Cosme19,Fujimoto19,Nagao19,Takasu20,Ozaki20} for details. 

\section{Details of the SU(3)DTWA simulation} \label{app:ppo}

First let us present the numerical sampling when the $x$-polarized state is chosen as our initial state. 
Generally speaking, the discrete-Wigner function representing such a superposed state exhibits negativity. 
To carry out an efficient numerical simulation, we take the following steps:

As the first step, we prepare a polarized down-spin state along $z$-axis at $t=-\pi \equiv t_0$
\begin{align}
|\Psi_0' \rangle = \bigotimes_{j=1}^{M}|S^{z}_{j}=-1\rangle. \label{eq: app: down-spin}
\end{align}
If we use the Wootters representation for the phase-point operator, the corresponding discrete-Wigner function is positive. 
Therefore, it is easy to sample randomized spins from the distribution. 

\begin{figure*}
\begin{center}
\includegraphics[width=180mm]{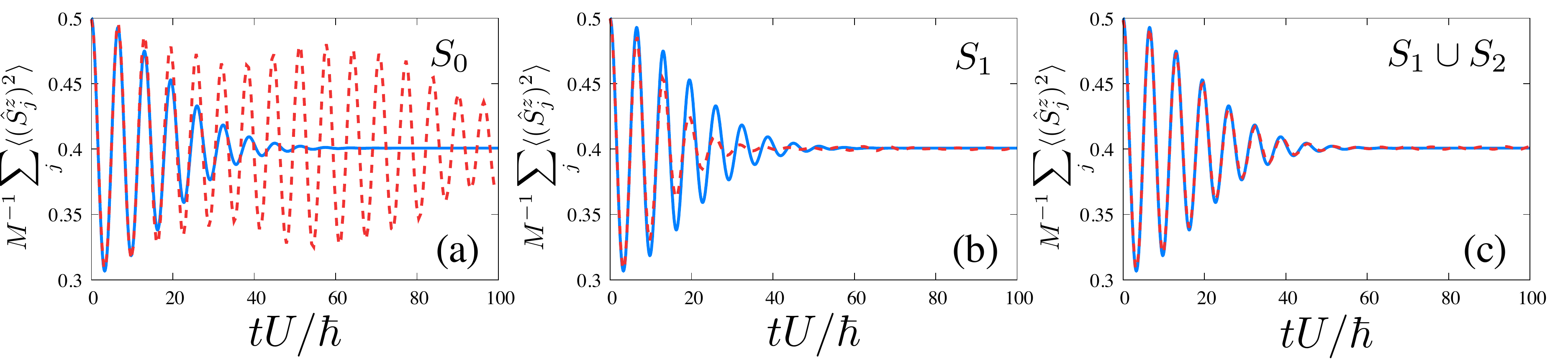}
\vspace{-5mm}
\caption{
SU(3)DTWA simulation for the fully connected spin-1 model for several statistical ensembles, namely (a) $S_{0}$, (b) $S_{1}$, and (c) $S_{1} \cup S_{2}$. 
The system size is $M=50$ and the interaction is $U/J=250$. 
The dashed lines are the results of the SU(3)DTWA.
The solid line represents the exact quantum dynamics.
The left ensemble $S_0$ corresponds to the Wootters representation.
}
\label{figure_appendix_02}
\end{center}
\end{figure*}

Then, we shine a global pulse such that it evolves $|\Psi_0' \rangle$ into the desired target state, i.e., the polarized state in the $x$-axis $|\Psi_0\rangle=\bigotimes_{j} |S^{x}_{j}=1\rangle$. 
Such a spin-flip process is designed via a unitary time evolution described by ${\hat U}_{\rm p}(t)=e^{-\frac{i}{\hbar}{\hat H}_{\rm p} (t-t_0)}$ with a Hamiltonian 
\begin{align}
{\hat H}_{\rm p} = \frac{1}{\sqrt{2}}\sum_{j=1}^{M}({\hat X}_{1}^{(j)}-{\hat X}_{3}^{(j)}).
\end{align}
If the unitary operation of the pulse is applied from $t=t_0$ to $t=0$, each spin state is locally flipped such that $|S^{z}_{j}=-1\rangle \rightarrow -|S^{x}_{j}=1\rangle$. 
The minus sign at the final state gives no effect on the expectation value. 
Notice that the time evolution governed by ${\hat U}_{\rm p}$ is exactly simulated by the SU(3)TWA because ${\hat H}_{\rm p}$ is linear in the phase-space variables. 
At $t=0$, the prepared random values of the spin configurations are expected to obey the Wigner distribution of $|\Psi_0\rangle=\bigotimes_{j} |S^{x}_{j}=1\rangle$.

Figure~\ref{figure_appendix_02}(a) displays the time evolution of the expectation value $M^{-1}\sum_{j}\langle ({\hat S}^{z}_{j})^2 \rangle$ for the Hamiltonian of the fully connected spin-1 model. 
The dashed line in Fig.~\ref{figure_appendix_02}(a) is calculated by using the Wootters representation $A^{(0)}_{\alpha}$ and following the above procedure. 
In what follows, we write $S_{0}$ as a statistical ensemble of the discretized phase-space variables sampled from the discrete-Wigner function for $A^{(0)}_{\alpha}$.
It is clear that the SU(3)DTWA with $S_0$ fails to reproduce the exact dynamics even though the Gaussian approach can do so. 
This consequence seems to be related to the fact that realizable configurations in $S_0$ are quite restricted compared with those belonging to the Gaussian distribution. 

To resolve this problem, we utilize a prescription in which we define a few other sets of the phase-point operator and make a statistical mixture of them as done in Ref.~\cite{Pucci16}. 
As a non-trivial example, we can construct the following two sets of phase-point operators instead of $A^{(0)}_{\alpha}$:
\begin{align}
A^{(1)}_{\alpha} &=
\begin{bmatrix}
\delta_{a_1,0}&\delta_{a_1,1}e^{-i\frac{2\pi a_2}{3}}&\delta_{a_1,2}e^{-i\frac{4\pi a_2}{3}}\\
\delta_{a_1,1}e^{i\frac{2\pi a_2}{3}}&\delta_{a_1,1}&\delta_{a_1,0}e^{-i\frac{2\pi a_2}{3}}\\
\delta_{a_1,2}e^{i\frac{4\pi a_2}{3}}&\delta_{a_1,0}e^{i\frac{2\pi a_2}{3}}&\delta_{a_1,2}
\end{bmatrix}, \\
A^{(2)}_{\alpha} &= 
\begin{bmatrix}
\delta_{a_1,0}&\delta_{a_1,0}e^{-i\frac{2\pi a_2}{3}}&\delta_{a_1,1}e^{-i\frac{4\pi a_2}{3}} \\
\delta_{a_1,0}e^{i\frac{2\pi a_2}{3}}&\delta_{a_1,1}&\delta_{a_1,2}e^{-i\frac{2\pi a_2}{3}} \\
\delta_{a_1,1}e^{i\frac{4\pi a_2}{3}}&\delta_{a_1,2}e^{i\frac{2\pi a_2}{3}}&\delta_{a_1,2} 
\end{bmatrix}.
\end{align}
It is straightforward to confirm that $A^{(1)}_{\alpha}$ and $A^{(2)}_{\alpha}$ satisfy the conditions in Sec.~\ref{sec: discrete wigner}.
This is obtained by simply exchanging ``$0$", ``$1$", and ``$2$" of the Kronecker deltas of the off-diagonal elements in $A^{(0)}_{\alpha}$. 
Notice that replacing $A^{(0)}_{\alpha}$ with $A^{(1)}_{\alpha}$ or $A^{(2)}_{\alpha}$ preserves the Wigner distribution for Eq.~(\ref{eq: app: down-spin}).
For the down-spin state (\ref{eq: app: down-spin}), $a_2$ is randomly distributed with probability $\frac{1}{3}$ and $a_1$ does not fluctuate, i.e., $a_1=2$.  
Therefore, we can make two independent statistical ensembles denoted by $S_{1}$ and $S_{2}$, respectively, in addition to $S_0$:
\begin{widetext}  
\begin{align}
(\sqrt{2}{\rm cos}\frac{2\pi a_2}{3},\sqrt{2}{\rm sin}\frac{2\pi a_2}{3},-1,0,0,-\sqrt{2}{\rm cos}\frac{2\pi a_2}{3},-\sqrt{2}{\rm sin}\frac{2\pi a_2}{3},-\frac{1}{\sqrt{3}}) &\in S_{0}, \\
(0,0,-1,2{\rm cos}\frac{4\pi a_2}{3},2{\rm sin}\frac{4\pi a_2}{3},0,0,-\frac{1}{\sqrt{3}}) &\in S_{1}, \\
(\sqrt{2}{\rm cos}\frac{2\pi a_2}{3},\sqrt{2}{\rm sin}\frac{2\pi a_2}{3},-1,0,0,\sqrt{2}{\rm cos}\frac{2\pi a_2}{3},\sqrt{2}{\rm sin}\frac{2\pi a_2}{3},-\frac{1}{\sqrt{3}}) &\in S_{2}.
\end{align}
\end{widetext}
As discussed in Ref.~\cite{Pucci16}, one can also make a statistical mixture such as $S_{0}\cup S_{1}$ and $S_{1}\cup S_{2}$. 
For example, $S_{1}\cup S_{2}$ means that a certain configuration in $S_{1}$ is realized with probability $\frac{1}{6}$ rather than $\frac{1}{3}$.

Figure~\ref{figure_appendix_02}(b) shows the SU(3)DTWA simulation corresponding to $S_{1}$. 
We observe that the saturated behavior for $t>40 \hbar/U$ is reproduced by this modification. 
Moreover, if we sample the randomized phase-space variables from the mixed ensemble $S_{1}\cup S_{2}$, we obtain the result of Fig.~\ref{figure_appendix_02}(c), which reasonably reproduces the exact quantum dynamics up to $tU/\hbar=100$.
For all the results for the $x$-polarized state $|\Psi_0\rangle=\bigotimes_{j} |S^{x}_{j}=1\rangle$ in the main text, we have used this mixture $S_{1}\cup S_{2}$ to generate randomized trajectories.

We come across the similar problem for the zero-magnetization state $\bigotimes_{j} |S^{z}_{j}=0\rangle$. 
Experiencing several non-trivial trials, we find that $S_1$ provides a better result than $S_0$, $S_2$, $S_{1}\cup S_{2}$, and other combinations. 
The main results in Sec.~\ref{sec:4} have been calculated for this single ensemble.
We may heuristically argue that simply adding a certain ensemble to $S_1$ does not necessarily improve the simulation results. 

It should be stressed that no procedure has been established so far to make an optimal sampling scheme for arbitrary states. 
Thus, for the intermediate use of our SU(3) discrete sampling approach, a proper choice of the sets of the phase-point operators and their statistical mixing, which is fixed through comparisons with exact computations for some exactly tractable cases, is always required.

\section{Note on the tomography sampling method and the statistical mixture method} \label{app: gdtwa}

We revisit here the discrete TWA sampling problem for the polarized down-spin state $|S_z=-1\rangle$ as treated in Appendix~\ref{app:ppo}, and give a remark associated with the tomography sampling method.
In this appendix, we ignore the spatial dependence, for simplicity. 

The down-spin state has the non-zero variance of ${\hat S}_{x}$, i.e., $\langle {\hat S}_{x}^2 \rangle = \langle {\hat X}_{1}^2 \rangle = 1/2$, because the state is not the eigenstate of ${\hat S}_{x}$.
If we use the tomography scheme for the phase-space sampling, it is found to successfully reproduce the exact moment in the phase-space representation.
Indeed, the corresponding probability distributions for the fluctuations of ${\hat X}_{1}$ are obtained as
\begin{align}
p_{1}^{(1)} = 1/4\;\;\; &\leftrightarrow \;\;\; \lambda^{(1)}_{1} = 1, \nonumber \\
p_{1}^{(2)} = 1/2\;\;\; &\leftrightarrow \;\;\; \lambda^{(2)}_{1} = 0, \nonumber \\
p_{1}^{(3)} = 1/4\;\;\; &\leftrightarrow \;\;\; \lambda^{(3)}_{1} = -1, \nonumber 
\end{align}
and result in $\langle {\hat X}_{1}^2 \rangle = \sum_{i=1}^{3}p_{1}^{(i)}(\lambda^{(i)}_{1})^2 = 1(1/4) + 0(1/2) + 1(1/4) = 1/2$.

The discrete sampling scheme based on the ensemble $S_1$ ($S_2$), however, fails to reproduce the moment in the phase-space representation.
In fact, the phase-space average $\langle {\hat X}_{1}^2 \rangle \rightarrow \overline{x_1^2}$ produces $0$ ($1$) as checked via direct computations. 
Hence, there is an underestimation (overestimation) of the quantum correlation in the classical ensemble generated by the naive phase-point-operator method. 
Note that, if in the beginning the squared operator ${\hat X}_{1}^2$ is linearized in the SU(3) matrices, and after that it is transformed to the phase-space quantities, the point-operator method accurately reproduces the moment.
The statistical mixture $S_1 \cup S_2$ that we have made in the previous appendix adequately averages the fluctuations of the classical variable belonging to each ensemble, and, as the consequence, produces the exact value of the moment as the phase-space average [namely, in this case, $(0+1)/2=1/2$].
This observation catches an underlying reason of the success of the DTWA simulation for the state $|S_x=1\rangle$, which is prepared after the unitary evolution of $|S_z=-1\rangle$ (see also Appendix~\ref{app:ppo}).
We expect that the tomography scheme will also give the adequate sampling for the simulation, but it is not explicitly implemented in this paper.
Thorough analyses about the connection between our DTWA scheme and the tomography method will be addressed elsewhere, which are beyond the central purpose of this work.

\section{Supplemental data for Sec.~\ref{sec:discussions}} \label{app: supplemental}

To visualize how the three-state truncation works in the parameter regime of the experiment, we numerically integrated the time-dependent Schr\"odinger equation for the 2D Bose-Hubbard Hamiltonian with $M=2^2=4$ and some values of $n_{\rm max}$.
Recall that $n_{\rm max}$ means the maximum occupation of each site.
The initial state of the following simulation is the unit-filling and homogeneous Mott-insulator state.

\begin{figure}
\begin{center}
\includegraphics[width=90mm]{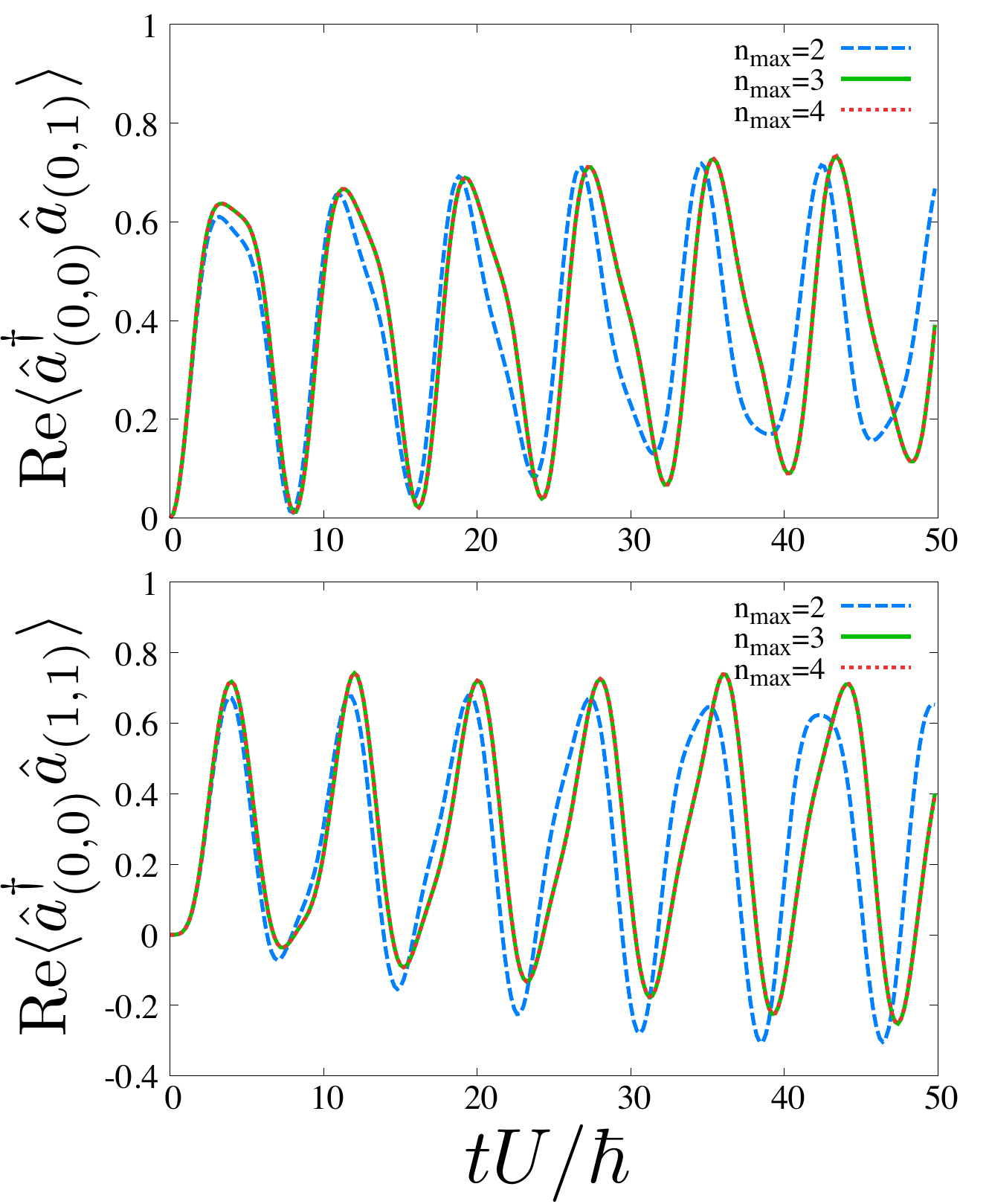}
\vspace{-3mm}
\caption{
Exact time evolution of the single-particle correlation function $\langle {\hat a}^{\dagger}_{j}{\hat a}_{j'} \rangle$ for a small size ($M=2^2=4$ sites).
The interaction is set to $U/J=20$.
The upper (lower) panel corresponds to ${\bm \varDelta}=(0,1)$ [${\bm \varDelta}=(1,1)$].
The blue dashed, green solid, and red dotted lines correspond to $n_{\rm max}=2$, $3$, and $4$, respectively.
We have imposed periodic boundary conditions on the system.
}
\label{fig: exact_smallsize}
\end{center}
\end{figure}

In Fig.~\ref{fig: exact_smallsize}, we show exact numerical results for the quench dynamics of the single-particle correlation function $\langle {\hat a}^{\dagger}_{j}{\hat a}_{j'} \rangle$.
The interaction during the time evolution is set to $U/J=20$, which is close to the actual value of the experiment, i.e., $U/J=19.6$.
The results for $n_{\rm max}=3$ (green solid line) and $n_{\rm max}=4$ (red dotted line) agree with each other, indicating that 4-particle occupations are completely suppressed at least until $t=50\hbar/U = 2.5 \hbar/J$.
Although the result for $n_{\rm max}=2$ (blue dashed line), which corresponds to the assumptions of the SU(3)TWA, fails to perfectly reproduce the result for $n_{\rm max}=3$, it captures very well the short-time evolution of the peak region of the correlations within $t \leq \hbar/J$.
Indeed, the peak region at early times agrees well with the one for $n_{\rm max}=3$ and the intensity of the correlation is close to the exact one.
As the system evolves in time, the deviation between the results for $n_{\rm max}=2$ and $3$ gradually gets significant.

%


\begin{thebibliography}{51}
\bibitem [{foo()}]{footnote}%
  \BibitemOpen
  \href@noop {} {\ }\bibinfo {note} {Under the periodic boundary conditions,
  $\varDelta_x = \text{min}(|x_{j}-x_{j'}|,|x_{j}-x_{j'}\pm L_x|$ and
  $\varDelta_y = \text{min}(|y_{j}-y_{j'}|,|y_{j}-y_{j'}\pm L_y|)$ are
  satisfied for a certain pair of $j$ and $j'$. Here we assume $\varDelta_x <
  L_x/2$ and $\varDelta_y < L_y/2$ ($M=L_xL_y$).}\BibitemShut {Stop}%
\end{thebibliography}
\end{document}